\documentclass[12pt]{article}

\usepackage{amssymb}
\usepackage{amsmath}
\usepackage{graphics,graphicx}

\newcommand{\singlespace}{\baselineskip=12pt \lineskiplimit=0pt \lineskip=0pt}
\newcommand{\beq}{\begin{equation}}
\newcommand{\eeq}{\end{equation}}
\newcommand{\lb}{\label}
\newcommand{\beqar}{\begin{eqnarray}}
\newcommand{\eeqar}{\end{eqnarray}}
\newcommand{\bit}{\begin{itemize}}
\newcommand{\eit}{\end{itemize}}
\newcommand{\barr}{\begin{array}}
\newcommand{\earr}{\end{array}}
\def\ds{\displaystyle}

\def\scalp{\mbox{\boldmath $\, \cdot \,$}}
\def\bob{{\, \underline{\overline{\otimes}} \,}}

\newcommand{\deriv}[2]{\frac{\partial #1}{\partial #2}}

\def\b0{\mbox{\boldmath $0$}}

\def\bA{\mbox{\boldmath $A$}}
\def\bB{\mbox{\boldmath $B$}}
\def\bC{\mbox{\boldmath $C$}}

\def\bP{\mbox{\boldmath $P$}}
\def\bQ{\mbox{\boldmath $Q$}}

\def\bS{\mbox{\boldmath $S$}}

\def\Id{\mbox{\boldmath $I$}}
\newcommand{\bsigma}{\mbox{\boldmath $\sigma$}}
\newcommand{\bepsilon}{\mbox{\boldmath $\epsilon$}}

\def\f0{\mbox{$\mathbb{O}$}}

\def\fE{\mbox{$\mathbb{E}$}}

\def\fG{\mbox{$\mathbb{G}$}}

\def\fP{\mbox{$\mathbb{P}$}}

\def\tr{\mbox{$\mathrm{tr}$}}

\def\dev{\mbox{$\mathrm{dev}$}}

\def\exp{\mbox{$\mathrm{exp}$}}
\def\cos{\mbox{$\mathrm{cos}$}}
\def\sin{\mbox{$\mathrm{sin}$}}
\def\tan{\mbox{$\mathrm{tan}$}}

\def\log{\mbox{$\mathrm{log}$}}

\def\Orth{\mbox{$\mathsf{Orth}$}}
\def\Orth+{\mbox{$\mathsf{Orth^+}$}}

\def\ACSB{{\it Am.\ Ceram.\ Soc.\ Bull.}\ }

\def\CMAME{{\it Comput.\ Method.\ Appl.\ M.}\ }

\def\GEOT{{\it G\'{e}otechnique}\ }

\def\IJMS{{\it Int.\ J.\ Mech.\ Sci.}\ }

\def\IJNME{{\it Int.\ J.\ Numer.\ Meth.\ Eng.}\ }

\def\IJSS{{\it Int.\ J.\ Solids Struct.}\ }
\def\JACS{{\it J.\ Am.\ Ceram.\ Soc.}\ }

\def\JEMT{{\it ASME J.\ Eng.\ Mater.\ Tech.}\ }

\def\JMPS{{\it J.\ Mech.\ Phys.\ Solids}\ }
\def\JSHA{{\it J.\ Soc.\ Hungarian Arch.}\ }

\def\MDPM{{\it Mod.\ Dev.\ Powder Metall.}\ }

\def\MOM{{\it Mech.\ Materials}\ }

\def\MRSB{{\it MRS Bull.}\ }

\def\PRSL{{\it Proc.\ R.\ Soc.\ Lond.}\ }

\def\ZAMP{{\it Z.\ Angew.\ Math.\ Phys.}\ }

\begin{document}

\title{An elastoplastic framework for granular materials 
becoming cohesive through mechanical densification. \\
Part I - small strain formulation}

\author{\\Andrea Piccolroaz, Davide Bigoni and Alessandro Gajo \\ \\
Dipartimento di Ingegneria Meccanica e \\ Strutturale,
Universit\`a di Trento, \\ Via Mesiano 77, I-38050 Trento, Italia\\
email: piccolroaz@ing.unitn.it, 
bigoni@ing.unitn.it, gajo@ing.unitn.it}

\date{August 2, 2004}

\maketitle

\begin{abstract}
\noindent
Mechanical densification of granular bodies is a process in which a loose material becomes 
increasingly cohesive as the applied pressure increases. 
A constitutive description of this process faces the formidable problem that granular and dense materials have 
completely different mechanical behaviours (nonlinear elastic properties, yield limit, plastic flow 
and hardening laws), 
which must both be, in a sense, included in the formulation. A treatment of this problem is provided here, 
so that a new phenomenological, elastoplastic constitutive model is 
formulated, calibrated by experimental data, implemented and tested, that is 
capable of describing the transition between granular and fully dense states of a given material. 
The formulation involves a novel use of elastoplastic coupling to describe the dependence of cohesion and
elastic properties on the plastic strain. The treatment falls within small strain theory, which is thought
to be appropriate in several situations; however, a 
generalization of the model to large strain is provided in Part II of this paper.
\end{abstract}

{\sl Keywords}: Elastoplasticity; Granular materials; Mechanical densification; Forming;
Ceramic Materials.

\thispagestyle{empty} 

\newpage

\section{Introduction}

\subsection{A premise and the central problem}

Cold powder compaction is a process in which granular materials are made cohesive
through mechanical densification. Subsequent sintering usually completes the treatment 
and yields the desired mechanical properties of the final piece.
Since this process permits an efficient production of 
parts ranging widely in size and shape to close tolerances (Reed, 1995), there is an evident related 
industrial interest. For instance, metallurgical (German, 1984), pharmaceutical (Lordi and Cuiti\~{n}o, 1997) and
forming of traditional (e.g. ceramic tiles and porcelain products) and structural (e.g. chip carriers, spark plugs,
cutting tools) ceramics represent common applications.

A crucial step in the above technique is the production of green bodies 
(i.e. the solids obtained after cold forming of ceramic powders) 
possessing enough cohesion to remain intact 
after mold ejection, being essentially free of macro defects, and handleable without failure in the subsequent
treatments. This is not an easy task and indeed defects of various types are always present in the greens 
(Deis and Lannutti, 1998; Ewsuk, 1997; Hausner and Kumar-Mal, 1982; 
Thompson, 1981b), negatively affecting local shrinkage during sintering. 
In particular, defects can be caused by the densification
process, which may involve highly inhomogeneous strain fields, or by mold ejection\footnote{Glass and Ewsuk (1997) 
classify several types of damage such as for instance 
end and ring capping, laminations, shape distortions, surface defects, vertical
cracks, and large pores.}.

In the present article a new elastoplastic {\it phenomenological} model is formulated, developed and implemented, which is
capable of describing 
the deformation and consequent gain in cohesion of granular {\it ceramic} 
materials subject to mechanical loading. The model is calibrated by experimental data relative to a 
ready-to-press alumina powder, and numerical predictions are compared to experiments performed on simple 
forming processes.
The proposed constitutive framework allows the possibility of simulating  
cold forming of pieces 
and to predict inhomogenities in cohesion, density, and elastic properties and residual stress 
distribution within the green, thus 
allowing {\it a rational design} of pieces.

Models aimed at achieving results similar to those addressed here have been developed for metal powders
(Ariffin et al. 1998; 
Brown and Abou-Chedid, 1994; Brown and Weber, 1988; Khoei and Lewis, 1999; 
Lewis and Khoei, 1998; Lewis et al. 1993; Oliver et al. 1996; 
Redanz, 1999; 2001; Redanz and Tvergaard, 2003; Sun and Kim, 1997)
and for ceramic granular materials
(Ahzi et al. 1993; Aydin et al. 1997a,b; Brandt and Nilsson, 1998; 1999; 
Ewsuk et al. 2001; Keller et al. 1998; Piccolroaz et al. 2002; Zipse, 1997). 
Due to the plasticity of grains, metal powders are 
essentially different from ceramics, but nevertheless we point out that both the above classes of models cannot describe 
the features considered in the present paper\footnote{The model 
employed by Piccolroaz et al. (2002) is simply a modified Cam-clay calibrated on experimental results 
relative to ceramic powders. It is similar to many other available in the literature and even if it can be considered 
perhaps sufficient for certain engineering purposes, it is far from realistic and does not address the major
difficulties in modelling the densification process.} 
. In particular, the central point is that 
\begin{quote}
{\it the process of mechanical compaction requires the description of the transition from a granular 
to a dense or even a fully dense state. Since granular materials are characterized by mechanical properties much 
different from those typical of dense solids, the constitutive modelling must describe a transition between two distinctly
different states of a material.} 
\end{quote}
The above constitutive description has been obtained for the first time in the present article by suggesting a 
novel use of elastoplastic coupling theory (Hueckel, 1976), introducing two micromechanically based 
hardening rules, and employing a recently proposed 
yield function (Bigoni and Piccolroaz, 2004). 

It may be also important to mention that, even if the emphasis is placed here on materials where 
cohesion increases with pressure, `vice-versa', we believe our constitutive framework to be sufficiently general 
to describe progressive decohesion due to mechanically-induced damage of quasibrittle materials 
(concrete, cemented sand, and rock), a problem investigated by Lagioia and Nova (1995).

Ceramic forming may involve deformations up to 50\% and even greater, so that 
the need for a large strain formulation could be advocated. However, numerical simulations (some of which 
will be presented later) show that 
at least the `gross' material behaviour is dominated 
by nonlinearities already occurring when deformations are small, so that the small strain formulation presented here 
should not be considered inappropriate. 
For completeness, nevertheless, a large strain formulation, requiring a new extension of the 
concept of elastoplastic coupling, is given in the Part II of this article (Piccolroaz et al. 2005).

\subsection{The densification of ceramic powders and the related difficulties in the constitutive modelling}

The compaction of a ceramic powder is a process essentially consisting of three phases: 
(I) granule sliding and rearrangement, (II) granule deformation, and (III) granule 
densification. For obvious reasons, a sharp differentiation between the three phases cannot be established, but 
qualitative morphologies of granular arrangements during Phases I and II are  
shown respectively in the central and lower parts of Fig.~\ref{fig01}, with respect to 
the ceramic powder employed in our study (392 Martoxid KMS-96, defined in Appendix A and shown in the upper part of 
the figure in its loose state).
\begin{figure}[!htb]
\begin{center}
\vspace*{3mm}
\includegraphics[width=6cm]{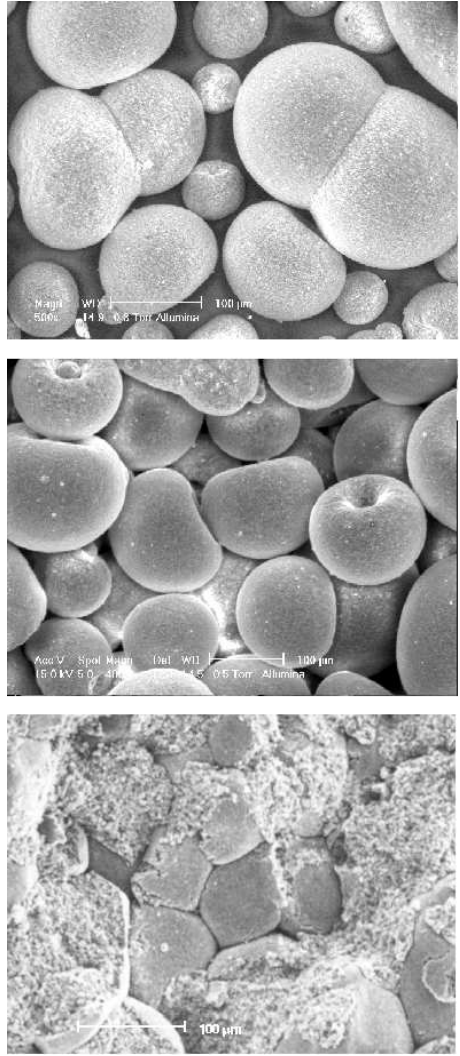}
\caption{\footnotesize Micrographs of M KMS-96 alumina powder. The loose state is shown in the 
upper part, while granule arrangements corresponding to Phases I and II are shown in the central and lower part of 
the figure.}
\label{fig01}
\end{center}
\end{figure}
 
\begin{figure}[!htb]
\begin{center}
\vspace*{3mm}
\includegraphics[width=7cm]{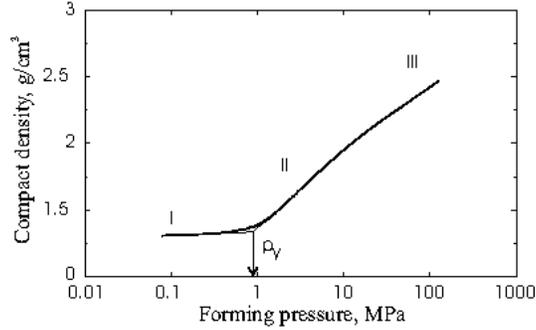}
\caption{\footnotesize Compaction diagram of M KMS-96 alumina powder. The breakpoint pressure 
(expressed in terms of axial stress during uniaxial strain), 
separating Phase I from Phase II,  
is denoted by $p_y$. A few identical experiments have been performed, providing almost coincident curves, one of 
which is reported here.}
\label{fig02}
\end{center}
\end{figure}

\begin{figure}[!htb]
\begin{center}
\vspace*{3mm}
\includegraphics[width=6.5cm]{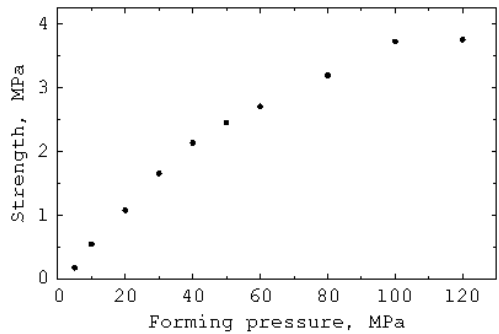}
\caption{\footnotesize Biaxial flexure strength (following ASTM F 394) of green tablets from  
M KMS-96 alumina powder,
as related to forming pressure under uniaxial strain in a cylindrical mold. 
Each point is the mean value taken over five tests, deviation was found
to be negligible.}
\label{fig03}
\end{center}
\end{figure}

The gain in cohesion starts with Phase II, which is 
marked by the so called `breakpoint pressure', conventionally denoted by a change
in inclination of the semi-logarithmic plot of density versus applied pressure (Fig.~\ref{fig02}, referred to ceramic 
tablets obtained by uniaxial 
pressing of 392 Martoxid KMS-96 alumina powder). 
Since the densification
process is often highly inhomogeneous, usually at least two phases coexist. Phase I always occurs in early volumetric 
deformation of granular materials, so that it has been thoroughly investigated for geomaterials and, 
as will be explained later,
is characterized by an elastic-plastic behaviour in which the elastic response is nonlinearly 
dependent on mean stress. 
In contrast to Phase I, the other phases have been much
less explored. The increase in cohesion\footnote{The degree of induced cohesion during mechanical 
densification depends on the 
applied pressure and is also strongly influenced by several factors, including  
stiffness, ductility and 
shape of the particles (Brown Abu Chedid, 1994). 
Experiments performed by D. Bigoni, G. Celotti, S. Guicciardi and A. Tampieri 
showed that penny-shaped particles produce a much higher increase in cohesion than spherical-shaped particles.} 
becomes already substantial during Phase II, as shown in Fig.~\ref{fig03}, where cohesion measured 
(on the same tablets employed for Fig.~\ref{fig02}) from biaxial flexure strength apparatus (defined by ASTM F 394) is reported versus the 
forming pressure (the axial compressive stress applied during uniaxial strain).

\begin{figure}[!htb]
\begin{center}
\vspace*{3mm}
\includegraphics[width=7cm]{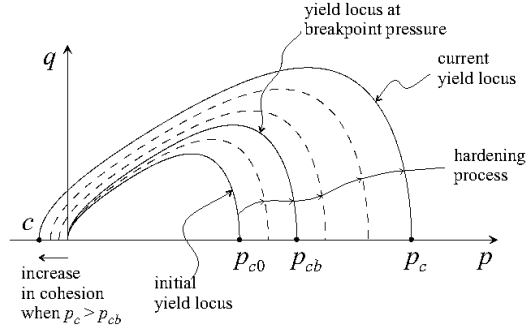}
\caption{\footnotesize Increase in cohesion as related to hardening.}
\label{fig04}
\end{center}
\end{figure}

The peculiar mechanism of variation in cohesion due to plastic deformation can be described by making recourse to the 
concept of hardening. In particular, we will assume that a yield function exists for a granular material, 
defining its elastic range, so 
that when the material is in the initial cohesionless state, the null stress state lies on the yield surface, 
Fig.~\ref{fig04}. 
Now, if the material is subject to increasing hydrostatic compression, after an initial (small) deformation in the 
elastic range, an early development of plastic deformation occurs from a virgin state, corresponding to Phase I 
compaction. In this phase, the increase in cohesion is limited and almost negligible. However, when the pressure 
reaches the breakpoint value, so that material enters Phase II, the gain in cohesion becomes crucially 
important. In conclusion, to describe this process, we may employ a hardening law leading to 
a yield surface evolution of the type sketched in 
Fig.~\ref{fig04}, where the yield surface shape distortion changes qualitatively, when the applied pressure $p$ 
exceeds the breakpoint pressure $p_{cb}$, expressed in terms of mean stress (with reversed sign)
\footnote{
The type of increase in cohesion could also qualitatively change after the Phase III of densification is 
entered, but since we do not possess enough experimental data relative to this behaviour (occurring however at 
very high pressures, not involved in the usual forming of ceramics), this is not accounted for in 
the modelling. We believe anyway that its consideration would be not difficult, once experimental results were 
made available.
}.

\begin{figure}[!htb]
\begin{center}
\vspace*{3mm}
\includegraphics[width=6cm]{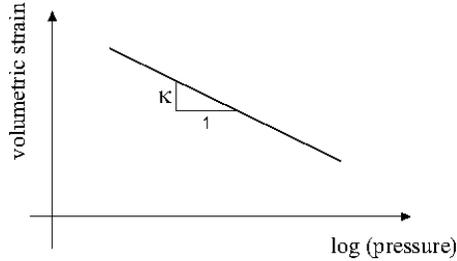}
\caption{\footnotesize Logarithmic elastic law relating the elastic bulk modulus $\kappa$ to pressure 
(mean stress taken positive when compressive).}
\label{fig05}
\end{center}
\end{figure}

At this level of description, one can get the impression that modelling the mechanism of cohesion increase during 
densification of granular materials could be pursued by simply
employing an appropriate hardening rule. However, {\it the elastic range of granular 
materials cannot be properly described by linear elasticity} so that the elastic response 
of the material changes, when the elastic range is modified during hardening. More in detail, during Phase I
of densification, studies relative to geomaterials demonstrate definitively 
that the elastic law 
relating the volumetric deformation to the applied 
mean pressure is logarithmic
\footnote{This law has reached the 
status of unchallenged law for fine grained soils, whereas it is strictly followed only during elastic 
unloading in coarse grained materials (Lambe and Whitman, 1969).}, as sketched in Fig.~\ref{fig05}.
Since the logarithmic law is simply not defined for a cohesionless material at null pressure, the increase in the 
cohesion of the material implies a modification to the elastic law, which becomes dependent on the plastic 
deformation, the physical quantity playing the role of the driving mechanism for densification. 
This means that {\it the elastic properties of the material must depend 
on the plastic deformation}, a feature that can be described by making recourse to the concept of elastoplastic 
coupling (Hueckel, 1976; Dougill, 1976).

In the following, we will introduce the three fundamental ingredients in the modelling of the densification 
processes, namely, (i) the yield function appropriate for the description of the behaviour of granular materials, 
(ii) the nonlinear elastic model, coupled to plasticity, (iii) two micromechanically-based hardening laws.

\subsection{Summary}

Part I of this paper is organized as follows. After a brief explanation of notation, the constitutive laws are 
introduced in Sect. \ref{costitut}, together with the calibration to experimental results. Numerical 
simulations and comparisons to experimental results relative to simple forming processed are presented in Sect. 
\ref{simulations}. Extension of the constitutive model to large strain is deferred to Part II of this paper 
(Piccolroaz et al. 2005).

\subsection{Notation}
A standard, intrinsic notation is used throughout the paper, where vectors and second-order tensors are denoted by 
bold (the latter capital) letters. The scalar product, the trace operator and the transpose are denoted by the 
usual symbols, namely,
\beq
\bA \scalp \bB = \tr \bA \bB^T,
\eeq
for every tensors $\bA$ and $\bB$.
The following invariants of stress $\bsigma$ will be used
\beq
\lb{invariantiterre}
p = -\frac{\tr \bsigma}{3},~~~ q = \sqrt{3 J_2},~~~ \theta = \frac{1}{3} \cos^{-1} \left(
\frac{3 \sqrt{3}}{2}
\frac{J_3}{J_2^{3/2}} \right),
\eeq
where $\theta \in [0, \pi/3]$ is the Lode's invariant and 
\beq
\lb{invariantisoliti}
J_2 = \frac{1}{2} \dev \bsigma \scalp \dev \bsigma,~~~
J_3 = \frac{1}{3} \tr \left(\dev \bsigma \right)^3,~~~
\dev \bsigma = \bsigma - \frac{\tr \bsigma}{3} \Id,
\eeq
in which $\dev \bsigma$ is the deviatoric stress and $\Id$ is the identity tensor. 

We will employ two tensorial products between second-order tensors $\bA$ and $\bB$, namely,
\beq
\left(\bA \otimes \bB\right)[\bC] = (\bB \scalp \bC)\bA,~~~ 
\left(\bA \bob \bB\right)[\bC] = \bA\frac{\bC + \bC^T}{2}\bB^T,
\eeq
so that $\Id \bob \Id$ becomes the symmetrizing fourth-order tensor, defined for every tensor $\bA$ as
$\Id \bob \Id[\bA] = (\bA + \bA^T) / 2$.

\section{The constitutive model} \lb{costitut}

\subsection{The yield function}

During cold mechanical compaction of ceramic powders, the shape of the yield surface evolves from that typical 
of granular material (for instance, that of the modified Cam-Clay model) to that characteristic of a ductile, dense material
(for instance, that of the Gurson model). To describe this process in terms of hardening, a suitably `deformable' 
surface is needed. This was found by Bigoni and Piccolroaz (2004); in particular, the yield function 
takes the form 
\beq
\lb{yieldfunction}
F(\bsigma, p_c, c)= f(p, p_c, c) + \frac{q}{g(\theta)},
\eeq
where $p_c$ and $c$ are the parameters that will be assumed to depend on plastic deformation and thus they will
define the hardening behaviour, $q$ is the deviatoric invariant 
(\ref{invariantiterre})$_2$, $f(p)$ is the function describing the dependence on the mean pressure $p$, 
eqn.~(\ref{invariantiterre})$_1$, assumed in the form
\beq
\lb{effedip}
f(p, p_c, c) =\left\{
\barr{ll}
-M p_c \sqrt{\left(\Phi - \Phi^m\right)\left[2 (1 - \alpha) \Phi + \alpha\right]} 
&~~~\mathit{if}\ \Phi\in[0,1],\\ [3 mm]
+ \infty &~~~\mathit{if}\ \Phi\notin[0,1],
\earr\right.
\eeq
in which 
\beq
\lb{fiegi}
\Phi = \frac{p + c}{p_c + c},
\eeq
and $g(\theta)$ describes dependence on the Lode's invariant $\theta$ defined by eqn.~(\ref{invariantiterre})$_3$
taken as
\beq \lb{gi}
g(\theta) = \frac{1}{\cos{\left[ \beta \frac{\pi}{6} - 
\frac{1}{3} \cos^{-1} \left(\gamma \cos{3 \theta}\right)\right]}}.
\eeq

The yield function described by eqn.~(\ref{yieldfunction})-(\ref{gi}) has been motivated and explained in great 
detail by Bigoni and Piccolroaz (2004), where the interested reader is remanded for details. We mention here that 
the yield surface corresponding to eqns.~(\ref{yieldfunction})-(\ref{gi}) is extremely versatile and remains 
convex in a broad range of values of parameters $M$, $p_c$, $c$, $m$, $\alpha$, $\beta$, $\gamma$. 

The yield function gradient $\bQ = \partial f/\partial \bsigma$, needed 
for the practical implementation of the model, is reported for completeness in Appendix B.

\subsubsection{Calibration of the parameter describing the yield function for alumina powder}

\begin{figure}[!htb]
\begin{center}
\vspace*{3mm}
\includegraphics[width=8cm]{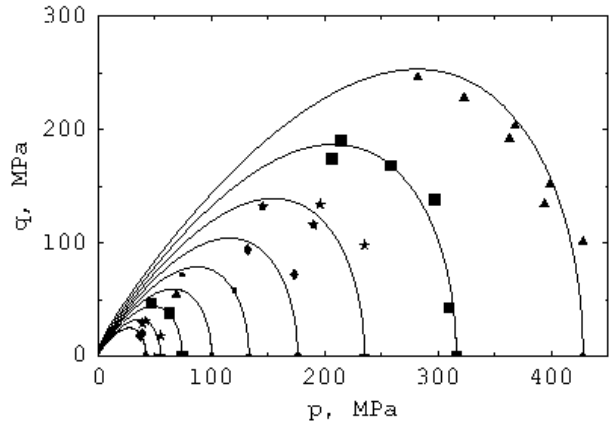}
\caption{\footnotesize Meridian sections of the employed yield surface fitted to the iso-density data 
for alumina powder (taken from Kim et al. 2002) at different 
levels of densification (increase in cohesion is not appreciable at the scale of the figure).}
\label{fig06}
\end{center}
\end{figure}

The problem now is to determine the yield surface parameters needed to describe the material considered here.
For this purpose, we note that 
$m$ and $\alpha$, defining the shape of the meridian section, 
can be calibrated employing the experimental iso-density data in the $(p,q)$ plane obtained from triaxial 
compression tests by Bonnefoy (as reported by Kim et al. 2002), under the assumption 
that iso-density curves correspond to yield surface sections (Kim et 
al. 2000; Kim et al. 2002). The values $m = 2$ and $\alpha = 0.1$ were found to provide the best fitting, as shown in 
Fig.~\ref{fig06}, where the data sets correspond to different levels of densification. 

To our knowledge, experimental data are not available to define the deviatoric section of the yield surface 
for alumina 
powder (and more in general for ceramic powders). Therefore, parameters $\beta$ and $\gamma$ 
have been calibrated on the basis 
of the value of the angle of internal friction, determined for our alumina powder from a standard 
geotechnical direct shear test apparatus\footnote{The
apparatus consists of a shear box containing the sample, which is split in the
middle height. When a normal force is applied, the horizontal force required to induce a
movement of the upper half of the sample with respect to the lower half is measured. This
test is useful for the evaluation of the friction angle of a granular material, like the
alumina powder in Phase I of densification. The samples were formed by carefully pouring
the ceramic powder within the shear box. Shearing was performed at a velocity of
0.2\,mm/min.} as follows. 

\begin{figure}[!htb]
\begin{center}
\vspace*{3mm}
\includegraphics[width=7.5cm]{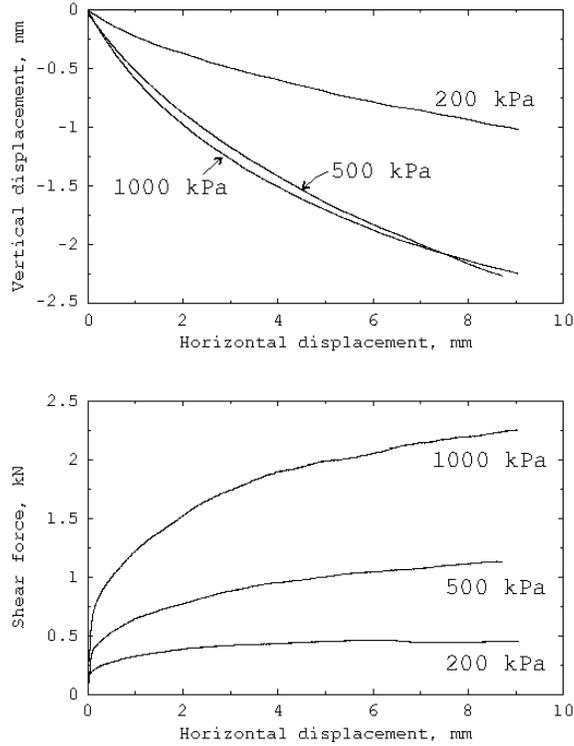}
\caption{Vertical (negative sign corresponds to contractant behaviour) vs.\ horizontal displacements (upper part) and 
shear force vs.\ horizontal displacement (lower part) of three samples of M KMS-96 alumina powder,
for different vertical pressures (200, 500 and 1000\,kPa).}
\label{fig07}
\end{center}
\end{figure}

The variation of the vertical displacement of the sample upper surface [a positive (negative) sign denotes a 
dilatant (contractant) behaviour] and
of the applied shear force during shearing is shown in Fig. \ref{fig07}. 

It can be observed from the figure that 
the tested samples exhibited the typical behaviour of a loose sand, with compressive volumetric
strains during shearing, without a softening phase. The fact
that the samples sheared at 500\,kPa and 1000\,kPa of vertical pressure have the same
volumetric strains is probably related to a slightly looser initial condition of the
former sample. It can be observed that, except for the test performed under a vertical
stress of 200\,kPa, the steady state condition typical of the critical state is not
reached and at the end of the test the strength and the volumetric strains of the samples
are still slightly increasing. This effect is more pronounced at larger applied vertical
pressures and is probably connected to the progressive deformation and rupture of the
grains constituting the alumina powder occurring during shearing even at low confining
pressures. This is consistent with the experimental evidences that very large shear
strains are necessary to reach the steady state in sands, when grain crushing occurs.

\begin{figure}[!htb]
\begin{center}
\vspace*{3mm}
\includegraphics[width=6cm]{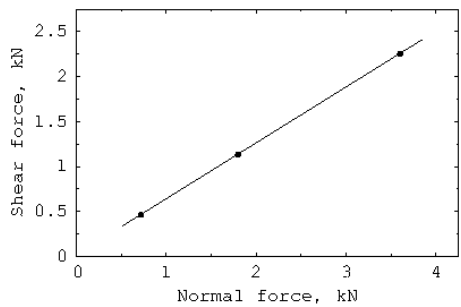}
\caption{\footnotesize Direct shear test on M KMS-96 alumina powder.}
\label{fig08}
\end{center}
\end{figure}

The maximum shear force reached at the end of the tests is plotted in Fig. \ref{fig08}
as a function of the applied vertical load. The results clearly lie on a straight line
and may be interpreted following the Coulomb-Mohr failure criterion, to yield a friction
angle approximately equal to 32$^\circ$. 
Making recourse now to the Coulomb-Mohr model, the ratio of the 
deviatoric section radius for triaxial extension, $g(0)$, to that for triaxial compression, $g(\pi / 3)$, is 
related to the angle of internal friction through
\beq
\lb{ratio}
\frac{g(0)}{g(\pi / 3)} = \frac{3 - \sin\, \phi}{3 + \sin\, \phi}.
\eeq
Eqn. (\ref{ratio}) is not enough to determine the two unknowns $\beta$ and $\gamma$. In the absence of further 
indications, parameter $\gamma$ was fixed to be equal to 0.9, providing a deviatoric section 
fairly close to the piecewise linear 
deviatoric section (corresponding to $\gamma = 1$). 
From this value and employing eqn.~(\ref{ratio}), $\beta = 0.19$ follows.

Parameter $M$, defining the pressure-sensitivity, has been calibrated making recourse to the concept of 
critical state, a peculiar state in which a granular material deforms at constant stress and constant volume. Since the 
critical state occurs at null plastic (and irreversible) volumetric strain rates, 
which (for the flow rule that will be specified later) is equivalent 
to $\tr \,\bQ = 0$, the expression for the yield function 
gradient eqn.~(\ref{grad})-(\ref{abc}) provides the following condition  
\beq
\lb{criticalstate}
2 (m + 1) (1 - \alpha) \Phi^m + m \alpha \Phi^{m-1} - 4 (1 - \alpha) \Phi - \alpha = 0,
\eeq
which identifies the critical state. 
This implicit relation can be solved numerically, as soon as values for $m$ and $\alpha$ have been selected, 
providing the value of $\Phi$ corresponding to the critical state, 
$$
\Phi^{\star} = \Phi^{\star} (m,\alpha).
$$ 
In our case $\Phi^{\star} = 0.658$ is obtained. The critical state point in the $(p,q)$ plane is therefore 
\beq
\lb{criticalstatepoint}
\left\{
\barr{l}
p^{\star} = (p_c + c) \Phi^{\star} - c, \\[5mm]
q^{\star} = g(\theta) M p_c \sqrt{(\Phi^{\star} - {\Phi^{\star}}^m) 
\left[ 2 (1 - \alpha) \Phi^{\star} + \alpha\right]},
\earr
\right.
\eeq
Eqns.~(\ref{criticalstatepoint}) define, through parameter $p_c$, the representation of the 
critical state line in the $(p,q)$ 
plane. Due to the fact that $c$ is a nonlinear function of $p_c$, the critical state line results mildly curved\footnote{
Note that the definition of critical state is extended here to include cohesive and ductile materials;
we assume for simplicity that cohesion does not change at critical state.}. 
Indeed, the line is straight only in the first phase of densification, 
as long as $c = 0$, and 
then deflects from linearity in the subsequent phase, when cohesion increases, approaching, after 
substantial plastic deformation and $c \sim c_{\infty}$, again a straight line with the same slope as the initial line. 
In other words, the final inclination of the critical state line, reached after substantial increase in cohesion, is
\beq
\lb{criticallineslope}
\frac{dq}{dp}=g(\theta) M \frac{\sqrt{(\Phi^{\star} - {\Phi^{\star}}^m) \left[ 2 (1 - \alpha) \Phi^{\star} + \alpha\right]}}
{\Phi^{\star}}.
\eeq
It can be observed from eqn. (\ref{criticalstatepoint}) that, since $c << p_c$ throughout the 
densification process, the effect of $c$ can be neglected, so that 
eqn.~(\ref{criticallineslope}) can be taken as the slope of the critical state line for 
the entire deformation process. With reference to 
the triaxial compression state, $\theta = \pi / 3$, this slope is related to the angle of internal friction 
through 
\beq
\frac{dq}{dp}=\frac{6\, \sin\, \phi}{3 - \sin\, \phi},
\eeq
which gives a value of $M$ equal to $1.1$.

\subsection{Elastoplastic coupling}

Dependence of elastic response of a material on plastic deformation for describing degradation of elastic 
properties was suggested independently by Hueckel (1976) for soils and Dougill (1976) for concrete. The 
model was later developed by Hueckel and Maier (1977), Capurso (1979), Maier and Hueckel (1979), and Bigoni and 
Hueckel (1991). We will develop the concept of elastoplastic coupling 
in the way suggested by Bigoni (2000), which yields a symmetric constitutive operator in the specific 
case of associative flow rule. This fact follows from different constitutive assumptions, in particular, it will 
be assumed here as in (Bigoni, 2000; Gajo et al. 2004) that the flow rule sets the so-called `irreversible' 
strain rate, which in the case of coupling is different from the plastic strain rate.

\begin{figure}[!htb]
\begin{center}
\vspace*{3mm}
\includegraphics[width=8cm]{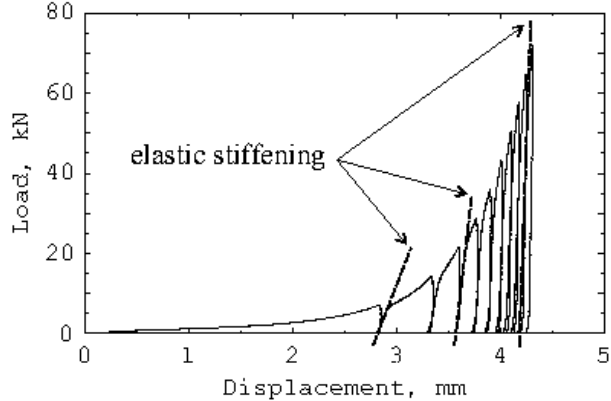}
\caption{\footnotesize Elastic stiffening during uniaxial deformation test (experimental results
on M KMS-96 alumina powder).}
\label{fig09}
\end{center}
\end{figure}

The necessity of elastoplastic coupling for modelling the densification process of granular materials considered here 
may be motivated ---as sketched in Fig.~\ref{fig09}--- by the observation that elastic unloading in a 
uniaxial deformation test shows a clear tendency toward a stiffening caused by the increase in cohesion. 
However, even if this effect would be disregarded in a first approximation, elastoplastic coupling would always 
be needed, for the reason mentioned in the Introduction, namely, to match the increase in cohesion 
with the nonlinear, logarithmic elastic model usually accepted for granular materials.

The basic concept of the elastoplastic coupling is that the elastic potential $\phi$ depends, in addition to the 
elastic, also on the plastic strain, so that 
\beq
\bsigma = \deriv{\phi \left( \bepsilon^e,\bepsilon^p \right)}{\bepsilon^e},
\eeq
where $\bepsilon^e$ and $\bepsilon^p$ are the elastic and plastic components of deformation, respectively (obeying
the usual addittive rule).

\subsection{Elastic potential for compaction during Phase I and early Phase II}

\begin{figure}[!htb]
\begin{center}
\vspace*{3mm}
\includegraphics[width=7cm]{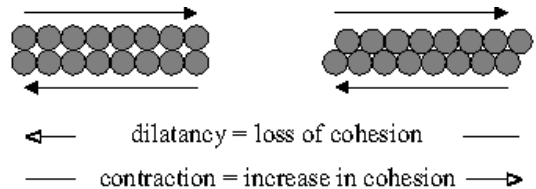}
\caption{\footnotesize The mechanism of increase and loss of cohesion visualized in terms of the Rowe model.}
\label{fig10}
\end{center}
\end{figure}

In the elastic range, granular cohesionless material 
obey the well-known logarithmic law sketched 
in Fig.~\ref{fig05}, relating the increment 
in the void ratio $\Delta e = e - e_0$ (measured with respect to an initial value $e_0$) to the current mean 
pressure
\beq
\lb{loglaw}
\Delta e^e = - \kappa\, \log\, \frac{p}{p_0},
\eeq
where the suffix $e$ remarks that we are referring to the elastic range, $p_0$ is the value of $p$ corresponding 
to the initial void ratio $e_0$, and $\kappa$ is the logarithmic bulk modulus. Eqn.~(\ref{loglaw}) is the 
starting point to obtain the nonlinear elastic potential employed in the Cam-clay model (Roscoe and Schofield, 
1963; Roscoe and Burland, 1968), which is particularly suitable for the description of cohesionless granular 
media, or, in other word, during Phase I of compaction\footnote{
The value of the logarithmic bulk modulus $\kappa$, which governs the elastic behaviour of the material in the 
first phase of densification, was deduced from experimental results for the alumina powder considered here
by measuring the slope of 
the curves obtained by loading and unloading 
in uniaxial strain tests. For this evaluation, we have assumed 
a constant ratio between the horizontal $\sigma_h$ and vertical $\sigma_v$ stresses equal to 0.47, as deduced 
from the formulae
\beq
\frac{\sigma_h}{\sigma_v} = 1 - \sin\, \phi, \nonumber
\eeq
which is usually employed for granular media (Jaky, 1944).}. 
However, our intention here is to describe the behaviour of materials which may increase (or decrease) 
cohesion as a function of the plastic deformation. Therefore, we have to introduce a modification in the elastic 
Cam-clay potential, first, to include a cohesion and, second, to make this dependent on plastic deformation. The 
easiest way to do this is to introduce a plastic-dependent cohesion in eqn.~(\ref{loglaw}), playing the role of a 
modification to the mean pressure
\beq
\lb{loglaw2}
\Delta e^e  = - \kappa\, \log\, \frac{p + c\left(\bepsilon^p\right)}{p_0 + c\left(\bepsilon^p\right)},
\eeq
where 
$c\left(\bepsilon^p\right)$ is the (positive) parameter describing the cohesion and depending on plastic 
deformation. In particular, the cohesion is assumed to depend on the {\it volumetric} component only of plastic 
deformation. This may be motivated by qualitative micromechanical considerations. Following, for instance, 
the Rowe (1962) model of a granular material, a shear deformation yields a decrease (increase) of cohesion 
when accompanied by dilatancy (contractivity), Fig.~\ref{fig10}.
Obviously, the relation (\ref{loglaw2}) is meaningful only in an {\it early stage of Phase II compaction}, since
unrealistic small values of void ratio are predicted at increasing $p$. However, it is important to realize
that the description of the initiation of Phase II is crucial, since the material goes from a cohesionless
state to a solid state.

We are in a position now to proceed with eqn.~(\ref{loglaw2}) in the way usually followed in the 
derivation of the Cam-clay model, i.e. in the case of 
eqn.~(\ref{loglaw}). Therefore, assuming incompressibility of grains, the volumetric elastic deformation is given by 
$(e-e_0)/(1+e_0)$, so that eqn.~(\ref{loglaw2}) defines a volumetric nonlinear elastic law, to be added to a 
linear elastic deviatoric constitutive equation. The resulting elastic potential is therefore 
\beq
\lb{elasticpotential}
\phi(\bepsilon^e,\bepsilon^p) = 
- \frac{\mu}{3} (\tr\, \bepsilon^e)^2 + c\, \tr\, \bepsilon^e  
+ \tilde{\kappa} (p_0 + c)\, \exp \left( - \frac{\tr\, \bepsilon^{e}}{\tilde{\kappa}} \right)
+ \mu\, \bepsilon^e \scalp \bepsilon^e ,
\eeq
where $\mu$ is the elastic shear modulus and $\tilde{\kappa} = \kappa / (1 + e_0)$. 

The elastic potential (\ref{elasticpotential}), 
with parameters $\mu$ and $\kappa$ taken independent of the plastic deformation, is certainly 
suitable to describe the behaviour of the material in Phase I of the compaction process, 
since here the
material is still granular and $c=0$, so that the model reduces to the celebrated Cam-clay model. Moreover, 
the elastic potential (\ref{elasticpotential}) can describe the initiation of Phase II of densification, where
the increase in cohesion starts to play a role. On the other hand, the potential (\ref{elasticpotential}) embodies
a nonlinear increase of elastic bulk modulus with mean pressure, a feature which is clearly unrealistic
during the late Phase II of deformation.

\subsection{Elastic potential for compaction during the late Phase II}

In the late Phase II, the material becomes more similar to a porous solid than to a granular body so that 
its properties progressively change. 
Experimental evidence of the mechanical behaviour of green's bodies formed at pressures 
corresponding to Phase II deformation is scarce. In particular, experiments by Zeuch et al. (2001) on alumina 
powder evidence that elastic properties become linear functions of forming pressure $p_c$. Therefore, an elastic
potential such as 
\beq
\lb{late}
\phi(\bepsilon^e,\bepsilon^p) =  \left(\frac{K}{2} - \frac{\mu}{3} \right)
\left( \tr\, \bepsilon^e  \right)^2 + \mu\, \bepsilon^e \scalp \bepsilon^e ,
\eeq
can be expected, where both $\mu$ and $K$ are linear functions of $p_c$. 
Though eqn. (\ref{late}) is very simple, we believe that the introduction of more complicated 
laws would be straightforward.

\subsection{Elastic potential for compaction during Phases I and II}

During deformation, the material behaviour changes from that described by potential (\ref{elasticpotential}) 
to that 
corresponding to potential (\ref{late}).
In order to describe this transition in the material behaviour, we modify the elastic potential 
(\ref{elasticpotential}) as follows
\beqar
\lb{elasticpotential2}
\lefteqn{\phi(\bepsilon^e,\bepsilon^p) = 
- \frac{\mu(d)}{3} (\tr\, \bepsilon^e)^2 + c\, \tr\, \bepsilon^e} \\ 
& & ~~~~
+ (p_0 + c) \left[ \left(d - \frac{1}{d}\right) \frac{(\tr\, \bepsilon^e)^2}{2 \tilde{\kappa}}
+ d^{1/n} \tilde{\kappa}\, \exp \left( - \frac{\tr\, \bepsilon^e}{d^{1/n} \tilde{\kappa}}  \right) \right] 
+ \mu(d)\, \bepsilon^e \scalp \bepsilon^e , \nonumber
\eeqar
where $d \geq 1$ is a `transition' parameter depending on the plastic volumetric strain 
through the forming pressure and 
governing the passage from logarithmic to
linear law of elastic bulk modulus. Note that for $d=1$ and in the limit $d \longrightarrow \infty$ the 
potentials (\ref{elasticpotential}) and (\ref{late}) are respectively recovered. Finally, $n \geq 1$ is a 
material constant defining the decay of the exponential term. Moreover, the 
elastic shear modulus $\mu$ is taken dependent on plastic volumetric strain through parameters $d$ and $c$ as follows
\beq
\lb{mu}
\mu(d) = \mu_0 + c \left( d - \frac{1}{d} \right) \mu_1,
\eeq
where $\mu_0$, and $\mu_1$ are positive material constants.

In conclusion, the nonlinear elastic stress/strain law may 
be obtained from eqn.~(\ref{elasticpotential2}) and results dependent on the plastic strain through $c$ and $d$
(the dependence is often omitted in the following for conciseness) 
\beqar
\lb{elasticlaw}
\lefteqn{\bsigma = 
\left\{ -\frac{2}{3} \mu\, \tr\, \bepsilon^{e} + c \right.} \\ 
& & ~~~~~~~
\left.
+ (p_0 + c) 
\left[ \left(d - \frac{1}{d}\right) \frac{\tr\, \bepsilon^{e}}{\tilde{\kappa}}
- \exp \left( -\frac{\tr\, \bepsilon^{e}}{d^{1/n} \tilde{\kappa}} \right) 
\right] \right\} \Id + 2 \mu\, \bepsilon^{e}. \nonumber
\eeqar

Taking the time derivative\footnote{Time is intended here as a loading parameter, while the 
material behaviour is assumed inviscid.} of (\ref{elasticlaw}), we get the rate equations
\beqar
\lb{ratelaw}
\lefteqn{\dot{\bsigma} = 
\fE[\dot{\bepsilon}^e] + \dot{c}
\left[
1 + \left(d - \frac{1}{d} \right) \frac{\tr\, \bepsilon^{e}}{\tilde{\kappa}} 
- \exp
\left(
-\frac{\tr\, \bepsilon^{e}}{d^{1/n} \tilde{\kappa}}
\right)
\right] \Id} \nonumber \\
& & ~~~~~
+ \dot{d}\, \frac{p_0 + c}{\tilde{\kappa}}\, \tr\, \bepsilon^{e}
\left[
1 + \frac{1}{d^2}
- \frac{1}{n d^{1+1/n}}\, 
\exp 
\left( -\frac{\tr\, \bepsilon^{e}}{d^{1/n} \tilde{\kappa}} \right)
\right] \Id \\
& & ~~~~~~~~~~~~~~~~~~~
+ \dot{\mu} \left( -\frac{2}{3}\, \tr\, \bepsilon^{e} \Id + 2 \bepsilon^{e} \right) , \nonumber
\eeqar
where $\dot{c}$, $\dot{d}$, and $\dot{\mu}$ arise from elastoplastic coupling ($\dot{c}=\dot{d}=\dot{\mu}=0$ in 
the usual uncoupled models) and the elastic fourth-order tensor $\fE$, together with its inverse $\fE^{-1}$ 
(restricted to the space of all symmetric tensors) are given by 
\beq
\ds \fE = \left[ -\frac{2}{3} \mu + K_t \right] \Id \otimes \Id + 2 \mu \Id \bob \Id, ~~~
\ds \fE^{-1} = \frac{2 \mu - 3 K_t}{18 \mu K_t} \Id \otimes \Id + \frac{1}{2 \mu} \Id \bob \Id,
\eeq
in which the tangent bulk modulus $K_t$ depends on the plastic deformation through $c$ and $d$ and on the elastic 
deformation in the following way
\beq
\lb{cappat}
K_t = 
\frac{p_0 + c}{\tilde{\kappa}} 
\left[ d - \frac{1}{d} + d^{-1/n} \exp \left( -\frac{\tr\, \bepsilon^{e}}{d^{1/n} \tilde{\kappa}} \right) \right].
\eeq

We need now to specify the particular dependence of the transition parameter $d$ on the 
forming pressure $p_c$. To this purpose, we note that,
assuming the existence of a saturation threshold $c_\infty$ for the value of the cohesion, 
the asymptotic behaviours of the bulk modulus $K_t$, eqn. (\ref{cappat}), and the 
shear modulus $\mu$, eqn. (\ref{mu}), as $d \to \infty$ are
\beq
\lb{asy}
K_t ~ \sim ~ \frac{p_0 + c_{\infty}}{\tilde{\kappa}} d ~~~\mathrm{and}~~~ \mu ~ \sim ~ c_{\infty} \mu_1 d,
\eeq
respectively.
Since experimental results by Zeuch et al. (2001) suggest that $\mu$ and $K_t$ become linear functions of 
$p_c$ for large 
values of forming pressure, inspired by (\ref{asy}), we assume for simplicity that 
parameter $d$ is a linear function of forming pressure 
$p_c$, for values of pressure superior to the breakpoint threshold $p_{cb}$, so that 
\beq
\lb{di}
d = 1 + B <p_c - p_{cb}>,
\eeq
where $B$ is a positive material parameter and 
the symbol $<>$ denotes the Macaulay brackets operator (defined for every scalar $\alpha$ as 
$<\alpha> = (\alpha +|\alpha|)/2$).

It has to be noted that the determination of parameters $B$, $\mu_1$ and $n$ is not easy. In principle,
parameters $B$ and $\mu_1$ can be obtained matching the asymptotic behaviour (\ref{asy}) with experiments
of the type performed by Zeuch et al. (2001). However, precise determination of elastic constants of green's bodies
is certainly difficult, moreover, $n$ needs also to be determined. We have estimated $B$, $\mu_1$ on the basis 
of the experiments by Zeuch et al. (2001), finding the values $B$=0.18 MPa$^{-1}$, $\mu_1$=64.

\begin{figure}[!htb]
\begin{center}
\vspace*{3mm}
\includegraphics[width=9cm]{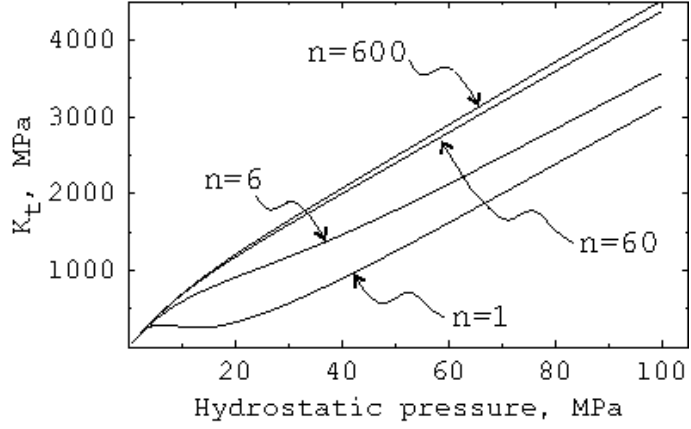}
\caption{\footnotesize Simulated evolution of tangent elastic bulk modulus during isotropic compression. Four values of $n$ are 
considered: $n = 1, 6, 60, 600$.}
\label{fig11}
\end{center}
\end{figure}

Regarding the constant $n$, we have plotted the evolution of the tangent elastic bulk modulus during a
hypothetical isotropic compression test at different values of $n$. This has been possible employing 
eqn. (\ref{cappat}) with $d$ given by eqn. (\ref{di}) 
and formula (\ref{ci}) for the cohesion $c$ that will be introduced later. 
In this way, $K_t$ depends on
the forming pressure $p_c$ (which has been taken coincident with $-\tr\,\bsigma/3$) 
and on $\tr\,\bepsilon$, which has been evaluated 
numerically solving the trace of eqn. (\ref{elasticlaw}). Results have been reported in Fig. \ref{fig11} for 
$n =$ $\{1, 6, 60, 600\}$,
from which we note that for $n$ = 1 unphysical behaviours appear, but these are not any more 
evident starting from $n$ = 6.
In the absence of experimental results, we have taken $n=6$ in our analyses. We note also that sensitivity 
to this parameter becomes very low for values of $n$ superior to 60, so that a refined determination of parameter $n$
would require {\it ad hoc} experimental investigation.

\subsection{Two micromechanically-based hardening laws}

In order to further develop eqn.~(\ref{ratelaw}), evolution laws for the hardening parameters representing the 
forming pressure $p_c$ and the cohesion $c$
are needed, providing the functional dependences of these parameters on the plastic deformation. 
At this point,  
recourse to micromechanical considerations and experimental evidence becomes necessary.

\subsubsection{The relation between the forming pressure $p_c$ and plastic volumetric strain}

Parameter $p_c$ is related to the plastic deformation and this 
relationship can be determined employing the micromechanical model proposed by Cooper and Eaton (1962). This 
takes into account the fundamental fact that compaction can be divided into the three phases mentioned in the 
introduction. Based on 
statistical micromechanics considerations and validated on several experimental results on ceramic powders, 
Cooper and Eaton (1962) provide a double-exponential law describing the first two phases of densification in 
terms of the relation between the plastic increment of void ratio $\Delta e^{p}$ and the pressure parameter $p_c$,
\beq
\lb{doppioexp}
-\frac{\Delta e^{p}}{ e_0 } = 
a_1\, \exp \left( -\frac{\Lambda_1}{p_c} \right) +
a_2\, \exp \left( -\frac{\Lambda_2}{p_c} \right),
\eeq
where $a_1$, $a_2$, $\Lambda_1$ and $\Lambda_2$ are material (positive) constants. In particular, coefficients 
$-e_0a_1$ and $-e_0a_2$ denote the increment of void ratio that would be achieved at infinite pressure by each of 
the two processes of densification, so that  $0 < a_1+a_2 \leq 1$. Coefficients $\Lambda_1$ and $\Lambda_2$, 
having the dimension of stress, indicate the magnitude of the pressure at which the particular process of 
deformation has the maximum probability density.

Assuming incompressibility of the grain constituents\footnote{
Ceramic powders are usually obtained through spray-drying and are formed by granules 
with dimensions ranging between 50 and 200 $\mu$m, coated with the binder system. 
Granules are itself aggregates of crystals having dimensions on the order of 1 $\mu$m. The crystals are
here assumed to be incompressible, while compressibility of granules is due to their internal voids.
}, the plastic volumetric deformation is related to the plastic void 
ratio increment according to the rule
\beq
\lb{deltaep}
\Delta e^{p} = (1+e_0)\, \tr\, \bepsilon^{p},
\eeq
we get from eqn.~(\ref{doppioexp}) 
\beq
\lb{doppioexp2}
\tr\, \bepsilon^{p} = 
- \tilde{a}_1\, \exp \left( -\frac{\Lambda_1}{p_c} \right) 
- \tilde{a}_2\, \exp \left( -\frac{\Lambda_2}{p_c} \right),
\eeq
where $\tilde{a}_i = e_0 a_i / (1 + e_0)$, $i=1,2$. 

\begin{figure}[!htb]
\begin{center}
\vspace*{3mm}
\includegraphics[width=9cm]{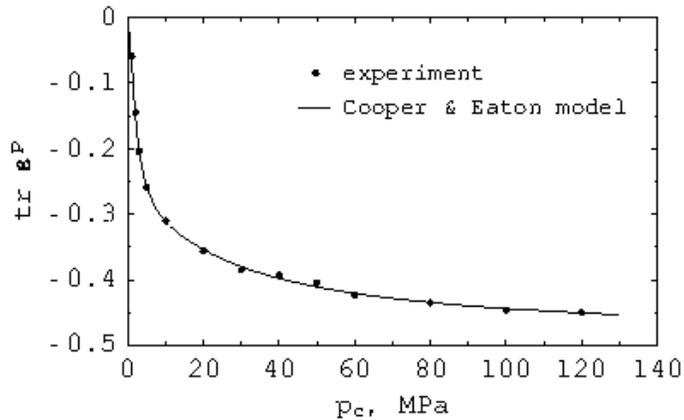}
\caption{\footnotesize Plastic volumetric strain as a function of the forming pressure $p_c$.
Experimental data relative to M KMS-96 alumina powder are compared with the model by Cooper and Eaton (1962) (solid line).}
\label{fig12}
\end{center}
\end{figure}

The hardening rule (\ref{doppioexp2}) is calibrated to describe uniaxial strain experiments, in which the 
permanent volumetric deformation has been measured at various forming pressure. In particular, taking the 
values $\Lambda_1 = 1.8$~MPa, $\Lambda_2 = 40$~MPa, $\tilde{a}_1 = 0.37$, 
$\tilde{a}_2 = 0.12$ gives the excellent interpolation presented in Fig.~\ref{fig12}, where the volumetric 
plastic strain $\tr \bepsilon^p$ is reported versus the hardening parameter $p_c$.
It may be noted that we restrict the attention to the double-exponential law (\ref{doppioexp2}) for simplicity, 
but it would be certainly not difficult to include more complicated relationships, which ---as suggested by Cooper and 
Eaton--- could include an arbitrary number of exponentials and therefore describe also the third phase of 
compaction behaviour.

Eqn.~(\ref{doppioexp2}) defines an implicit relation between 
the plastic deformation $\tr \, \bepsilon^p$ and the forming pressure $p_c$, 
which becomes explicit in terms of rates. In particular, the rate of 
eqn.~(\ref{doppioexp2}) gives
\beq
\lb{dotpc}
\dot{p}_c = 
- \frac{p_c^2}{\tilde{a}_1 \Lambda_1\, \exp \left(\ds{- \frac{\Lambda_1}{p_c}} \right)
+ \tilde{a}_2 \Lambda_2\, \exp \left(\ds{- \frac{\Lambda_2}{p_c}} \right)}\, \tr\, \dot{\bepsilon}^{p},
\eeq
providing the first hardening rule.

\subsubsection{Modelling the increase of cohesion}

\begin{figure}[!htb]
\begin{center}
\vspace*{3mm}
\includegraphics[width=9cm]{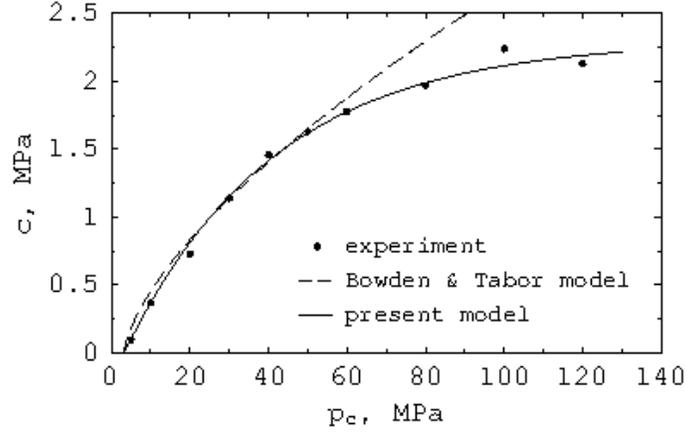}
\caption{\footnotesize Variation of the cohesion $c$ as a function of the forming pressure $p_c$. Experimental
results relative to M KMS-96 alumina powder are compared to our model (\ref{ci}) (solid line) and to that by Bowden and Tabor (1950) (dashed line).}
\label{fig13}
\end{center}
\end{figure}

Concerning the dependence of cohesion $c$ on the forming pressure $p_c$, we could recourse to models of adhesion
between particles. We note that the celebrated JKR model (Johnson et al. 1971) (and also 
variants like for instance the DMT model) is not applicable
in our case, since the ceramic granules considered here are highly plastic and the adhesion force cannot be treated as
independent of the granule deformation. It seems more appropriate to recourse to the Bowden and Tabor (1950) model for
adhesion between surfaces in contact. In this model, the cohesion depends on the real contact area and therefore 
on the normal pressure. Assuming a Herzian contact between spheres and that the cohesion $c$ is a linear function 
of the contact area, the following relationship 
\beq
\lb{tabor}
c \sim p^{2/3},
\eeq
is found, in which $p$ is the contact pressure. The major concern with condition (\ref{tabor}) 
is that it does not predict a limit for the increase of adhesion with pressure, which is a clear experimental
evidence. More in detail, our experimental results reported in Fig.~\ref{fig03} can be manipulated employing our 
model and expressed in terms of cohesion $c$ versus the forming pressure $p_c$. The results are shown in 
Fig.~\ref{fig13}, where interpolation using the Bowden and Tabor approximation (\ref{tabor}) is also reported 
(dashed). The solid line in the figure is obtained employing the following law
\beq
\lb{ci}
c = c_{\infty} \left[ 1- \exp \left(-\Gamma <p_c - p_{cb}> \right) \right],
\eeq
where $p_{cb}$ is the breakpoint pressure, $c_{\infty}$ and $\Gamma$ are two positive 
material parameters, the former 
defining the limit value of cohesion reached after substantial plastic deformation, the latter related 
to the 
`velocity of growth' of cohesion. The values $\Gamma = 0.026$~MPa$^{-1}$, $c_\infty = 2.3$~MPa, 
and $p_{cb} = 3.2$~MPa have been found to provide an excellent interpolation to experimental data.
Note that the Macaulay bracket is needed since the cohesion is null before the threshold value
defined by the breakpoint pressure is attained.

Taking now the rate of eqn. (\ref{ci}) and using (\ref{dotpc}) yields
\beq
\lb{dotci}
\dot{c} = \xi_2\, \tr\, \dot{\bepsilon}^{p},
\eeq
where
\beq
\lb{xi2}
\xi_2 =  
- \frac{c_{\infty} \Gamma\, H(p_c - p_{cb})\, \exp \left[ -\Gamma ( p_c - p_{cb}) \right] p_c^2}
{\tilde{a}_1 \Lambda_1\, \exp \left(\ds{- \frac{\Lambda_1}{p_c}} \right)
+ \tilde{a}_2 \Lambda_2\, \exp \left(\ds{- \frac{\Lambda_2}{p_c}} \right)},
\eeq
providing the second hardening rule. Note that symbol $H$ in eqn. (\ref{xi2}) denotes the Heaviside step function 
(defined for every scalar $\alpha$ as $H(\alpha) = 1$, if $\alpha \geq 0$, $H(\alpha) = 0$ otherwise).

\subsection{The elastoplastic coupling in rate form}

Taking the time derivative of eqns. (\ref{di}) and (\ref{mu}) and considering eqn. (\ref{dotpc}) we get the 
dependence of elastic parameters on plastic strain in rate form
\beq
\lb{dotdimu}
\dot{d} = \xi_3\, \tr\, \dot{\bepsilon}^{p}, 
~~~\mathrm{and}~~~ \dot{\mu} = \xi_4\, \tr\, \dot{\bepsilon}^{p},
\eeq
where
\beq
\lb{xi3}
\xi_3 = 
- \frac{B\, H(p_c - p_{cb})\, p_c^2}{\tilde{a}_1 \Lambda_1\, \exp \left(\ds{- \frac{\Lambda_1}{p_c}} \right)
+ \tilde{a}_2 \Lambda_2\, \exp \left(\ds{- \frac{\Lambda_2}{p_c}} \right)},
\eeq
\beq
\lb{xi4}
\xi_4 = 
\left( d - \frac{1}{d} \right) \mu_1 \xi_2 + c \left( 1 + \frac{1}{d^2} \right) \mu_1 \xi_3 . \nonumber
\eeq

Employing the evolution equations (\ref{dotci}) and (\ref{dotdimu}) into 
the stress rate equations (\ref{ratelaw}), we may write 
\beq
\lb{ratelaw2}
\dot{\bsigma} = 
\fE[\dot{\bepsilon}^e]+\fP[\dot{\bepsilon}^p],
\eeq
where the fourth-order tensor $\fP$ defines the contribution of the elastoplastic coupling, in the sense that 
$\fP$ is null in the usual, uncoupled plasticity and is defined as
\beq
\lb{pi}
\fP = \xi_5 \Id \otimes \Id + 2 \xi_4 \bepsilon^e \otimes \Id,
\eeq
in which
\beqar
\lefteqn{\xi_5 = - \frac{2}{3} \xi_4\, \tr\, \bepsilon^e 
+ \xi_2 
\left[
1 + \left(d - \frac{1}{d} \right) \frac{\tr\, \bepsilon^{e}}{\tilde{\kappa}} 
- \exp
\left(
-\frac{\tr\, \bepsilon^{e}}{d^{1/n} \tilde{\kappa}}
\right)
\right]} \\
& & ~~~~~~~~~
+ \xi_3\, \frac{p_0 + c}{\tilde{\kappa}}\, \tr\, \bepsilon^{e}
\left[
1 + \frac{1}{d^2}
- \frac{1}{n d^{1+1/n}}\, \exp 
\left(-\frac{\tr\, \bepsilon^{e}}{d^{1/n} \tilde{\kappa}}\right)
\right], \nonumber
\eeqar
when elastoplastic coupling occurs.

Introducing now the `irreversible' (in an infinitesimal stress cycle) strain rate
\beq
\lb{irrevrate}
\dot{\bepsilon}^i = \fG[\dot{\bepsilon}^p],
\eeq
in which
\beq
\lb{gig}
\fG = 
\Id \bob \Id - \fE^{-1} \fP =
\Id \bob \Id + \xi_6 \Id \otimes \Id + \xi_7 \bepsilon^e \otimes \Id,
\eeq
and 
\beq
\xi_6 = 
- \frac{\xi_5}{3 K_t} - \frac{2 \mu - 3 K_t}{9 \mu K_t} \xi_4\, \tr\, \bepsilon^e, ~~~ 
\xi_7 = 
- \frac{\xi_4}{\mu},
\eeq
we may transform the rate equation (\ref{ratelaw2}) into the equivalent form
\beq
\dot{\bsigma} = 
\fE[\dot{\bepsilon}]-\fE[\dot{\bepsilon}^i] .
\eeq

Since tensor $\fG$ is assumed positive definite, implying that 
\beq
\dot{\bepsilon}^i \scalp \dot{\bepsilon}^p > 0,
\eeq
the inverse of $\fG$ is given by
\beq
\lb{giinv}
\fG^{-1} = 
\Id \bob \Id + \xi_8 \Id \otimes \Id + \xi_9 \bepsilon^e \otimes \Id, 
\eeq
where
\beq
\xi_8 = 
- \frac{\xi_6}{1 + 3 \xi_6 + \xi_7\, \tr\, \bepsilon^e}, ~~~ 
\xi_9 = 
- \frac{\xi_7}{1 + 3 \xi_6 + \xi_7\, \tr\, \bepsilon^e}.
\eeq

\begin{figure}[!htb]
\begin{center}
\vspace*{3mm}
\includegraphics[width=6cm]{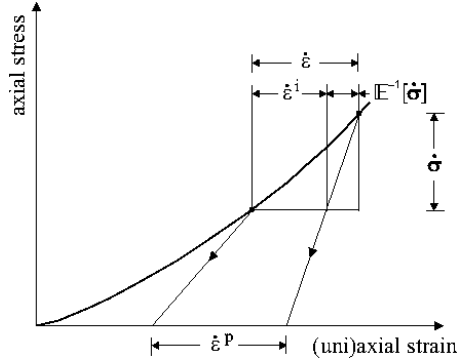}
\caption{\footnotesize Irreversible, and plastic rate deformations, 
with reference to volumetric deformation.}
\label{fig14}
\end{center}
\end{figure}

The irreversible deformation rate $\dot{\bepsilon}^i$ defined by eqn.~(\ref{irrevrate}) is the rate of 
deformation which is not recovered in an infinitesimal stress cycle. This should not be confused with the plastic 
deformation rate $\dot{\bepsilon}^p$, which can be only detected upon unloading at zero stress. This concept is 
illustrated in Fig.~\ref{fig14} with reference to a hypothetical volumetric stress/strain law, where 
the rate deformation at unloading $\fE^{-1}[\dot{\bsigma}]$ is also indicated. 
Due to the plastic increase in elastic stiffness, we note that (\ref{ratelaw}) implies that the plastic 
$\dot{\bepsilon}^p$ and inelastic $\dot{\bepsilon}^i$ rates are different.

\subsection{Flow rule}
A crucial point is now the definition of the flow rule, that following 
Bigoni (2000) is assumed to govern the irreversible strain rate (instead of the plastic, see Gajo et al. 2004 for
a discussion on this issue)
\beq
\lb{flowrule}
\dot{\bepsilon}^{i}=\dot{\lambda} \bP.
\eeq
Experimental evidence for granular material supports the use of a 
deviatoric associative flow rule, which is adopted here. Therefore, 
tensor $\bP$ is related to the yield function gradient $\bQ$ through 
\beq
\lb{flowmod}
\bP = \bQ - \frac{\epsilon \, (1 - \Phi)\, \tr\, \bQ  }{3}\, \Id,~~~ 0 \leq \epsilon \leq 1,
\eeq
where $\epsilon$ is a parameter governing the entity of volumetric nonassociativity, so that $\epsilon = 0$ gives 
the associative flow rule. All indirect evidences point to flow rule nonassociativity for ceramic powders, so that 
we feel that eqn. (\ref{flowmod}) is appropriate; however, experimental evidence for 
alumina powder is not available and associativity will be assumed for simplicity in the following numerical 
simulation.

The rate constitutive equations can now be obtained via Prager's consistency, so that $\dot{F}= 0$ during plastic 
deformation. Imposing this condition suggests the following definition of hardening modulus
\beq
\lb{hardeningmod}
h = -\frac{1}{\dot{\lambda}}
\left(
\deriv{F}{p_c} \dot{p}_c + \deriv{F}{c} \dot{c}
\right),
\eeq
which is positive in the case of hardening, negative for softening and null for ideally plastic behaviour. The 
derivatives of $F$ with respect to the hardening parameters $p_c$ and $c$ appearing in eqn.~(\ref{hardeningmod}) 
are given by
\beqar
\lefteqn{\deriv{F}{p_c} = -M\sqrt{\left( \Phi - \Phi^m \right)[2(1-\alpha) \Phi + \alpha]}} \\
& & 
+M \frac{p_c(p+c)}{(p_c+c)^2}
\frac{\left(1-m\Phi^{m-1}\right)\left[2(1-\alpha)\Phi+\alpha\right]+2(1-\alpha)\left(\Phi-\Phi^m\right)}
{2\sqrt{\left(\Phi-\Phi^m\right)\left[2(1-\alpha)\Phi+\alpha\right]}}, \nonumber
\eeqar
and
\beqar
\lefteqn{\deriv{F}{c} = 
-M \frac{p_c(p_c-p)}{(p_c+c)^2} \cdot} \\
& & ~~~~~~~~~~
\cdot \frac{\left(1-m\Phi^{m-1}\right)\left[2(1-\alpha)\Phi+\alpha\right]+2(1-\alpha)\left(\Phi-\Phi^m\right)}
{2\sqrt{\left(\Phi-\Phi^m\right)\left[2(1-\alpha)\Phi+\alpha\right]}}. \nonumber
\eeqar

\subsection{The final rate equations}

Employing definition (\ref{hardeningmod}) into Prager's consistency yields the elastoplastic rate equations
\beq
\dot{\bsigma} = 
\left\{ 
\barr{lll}
\ds
\fE[\dot{\bepsilon}] - \frac {<{\bf{Q}} \scalp \fE[\dot{\bepsilon}]> }{h + \bQ \scalp \fE[\bP]} \, \fE[{\bf{P}}] & ~~~
\mathrm{if}~~ F(\bsigma, p_c, c) = 0, \\[2mm] 
\fE[\dot{\bepsilon}] & ~~~
\mathrm{if}~~ F(\bsigma, p_c, c) < 0.
\earr 
\right.
\eeq
It may be noted that the elastoplastic tangent operator becomes symmetric in the specific case of the associative 
flow rule, $\bP = \bQ$.

\section{Numerical simulations}
\lb{simulations}

The proposed constitutive model was implemented into UMAT, the subroutine available within the commercial finite 
element code 
ABAQUS (Ver. 6.3; Hibbitt, Karlsson \& Sorensen, 2002,  Pawtucket, RI, USA). The employed numerical integration scheme 
was the so-called `cutting-plane algorithm', proposed by Simo and Ortiz (1985), 
Ortiz and Simo (1986), Simo and Huges (1987). A full Newton-Rapson scheme has been employed for 
the solution of the nonlinear finite element problem.
Parameters of the models employed for the simulations are
summarized in Table~\ref{values}.

 \begin{table}[!htb]
 \footnotesize
 \begin{center}
 \caption{Values of material parameters estimated from experiments for alumina 
powder 392 Martoxid KMS-96.}
 \label{values}
 \begin{tabular}{||c||}
 \hline
 \hline 
 Material parameters defining the yield surface                                           \\
 $m = 2$, $\alpha = 0.1$, $\beta = 0.19$, $\gamma = 0.9$, $M = 1.1$                       \\ 
 \hline
 Elastic logarithmic bulk modulus $\kappa = 0.04$                                         \\
 \hline 
 Material parameters defining the hardening rule (\ref{doppioexp2})                       \\
 $\Lambda_1 = 1.8$~MPa, $\Lambda_2 = 40$~MPa, $\tilde{a}_1 = 0.37$, $\tilde{a}_2 = 0.12$  \\
 \hline 
 Material parameters defining the hardening rule (\ref{ci})                               \\
 $\Gamma = 0.026$~MPa$^{-1}$, $c_{\infty} = 2.3$~MPa, $p_{cb} = 3.2$~MPa                  \\
 \hline 
 Material parameters defining coupling rules (\ref{di}) and (\ref{mu})                     \\ 
 $B = 0.18$~MPa$^{-1}$, $n = 6$, $\mu_0 = 1$~MPa, $\mu_1 = 64$                            \\
 \hline 
 Material parameter defining the flow rule $\epsilon = 0$                                 \\
 \hline 
 \hline
 \end{tabular}
 \end{center}
 \end{table}

\subsection{Forming of tablets}

\begin{figure}[!htb]
\begin{center}
\vspace*{3mm}
\includegraphics[width=15cm]{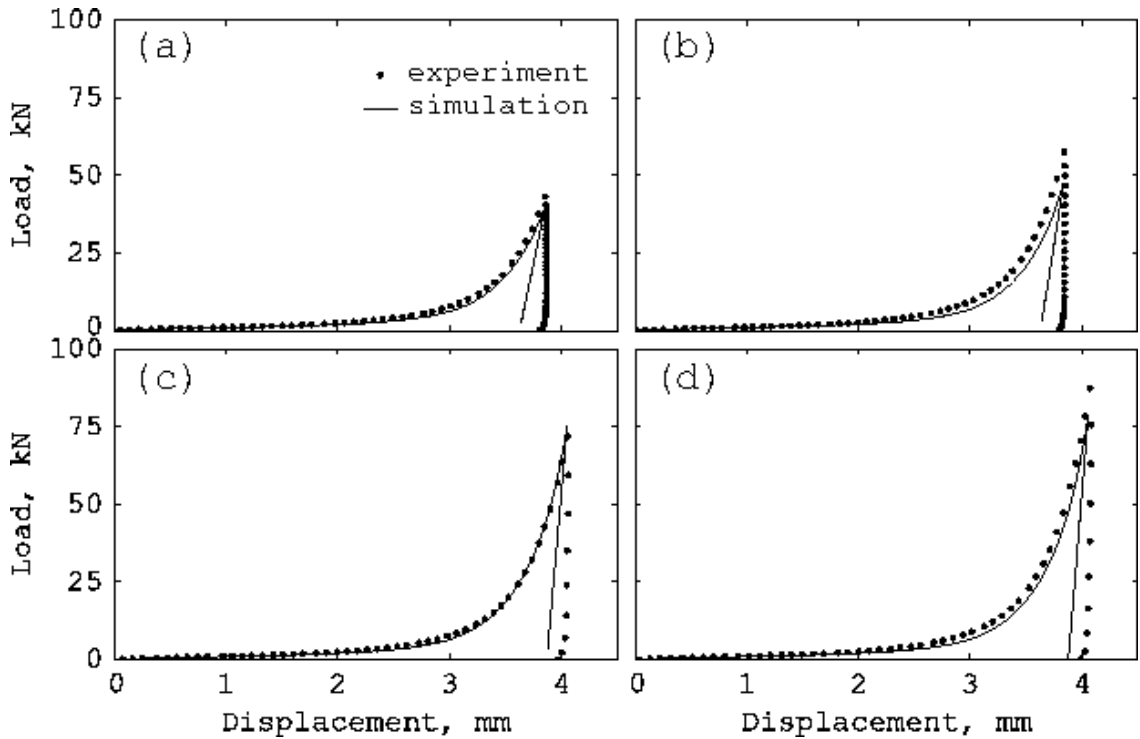}
\caption{\footnotesize Experimental (data relative to M KMS-96 alumina powder) 
and simulated load vs.\ displacement curves for tablets formed at
various final pressures: 60 (a), 80 (b), 100 (c), and 120 MPa (d). Different values of $n$ have been considered.}
\label{fig15}
\end{center}
\end{figure}

Simulations of uniaxial deformation of cylindrical samples, or, in other words, forming of 
tablets, are reported in Fig.~\ref{fig15}, together with our experimental results, 
marked by spots.  
In the figure the applied vertical load (in kN) is reported versus the vertical displacement 
(in mm) and different forming pressures, equal to $\{60, 80, 100, 120\}$ MPa, have been considered.
The fact that the experiments are correctly simulated employing the finite element discretisation
should be considered as a succesful feedback on the `consistency' of the model and its implementation.
Moreover, we note that the model describes the progressive increase in elastic stiffness (visible at unloading) 
in qualitative and quantitative agreement with experimental data.

\subsection{Forming of a simple ceramic piece}

\begin{figure}[!htb]
\begin{center}
\vspace*{3mm}
\includegraphics[width=7cm]{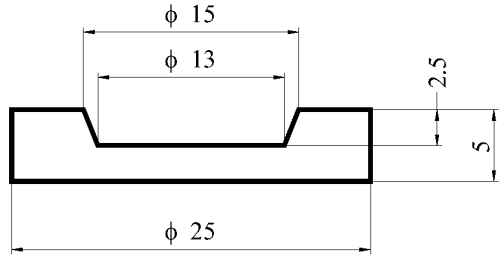}
\caption{\footnotesize Geometry of the formed piece (dimensions in mm).}
\label{fig16}
\end{center}
\end{figure}

Numerical simulations were performed to describe forming of the (axisymmetric) piece 
geometrically described in Fig. \ref{fig16} and shown as a green body after forming in
Figs.~\ref{fig16}. 
\begin{figure}[!htb]
\begin{center}
\vspace*{3mm}
\includegraphics[width=6cm]{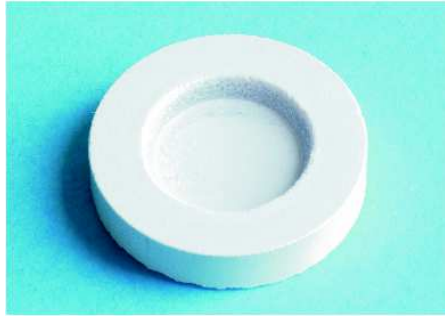}
\caption{\footnotesize Photograph of the formed green piece (5~g of M KMS-96 alumina powder has been used and 
a final mean pressure of 100 MPa has been reached).}
\label{fig17}
\end{center}
\end{figure}
In particular, four pieces have been formed at a final mean pressure of 100~MPa, starting from 5~g 
of the M KMS-96 alumina powder employed in all our experiments. 
The axisymmetric mesh, employing 4-node elements (CAX4),  
used in the simulations is shown in Fig.~\ref{fig18}. 
\begin{figure}[!htb]
\begin{center}
\vspace*{3mm}
\includegraphics[width=8cm]{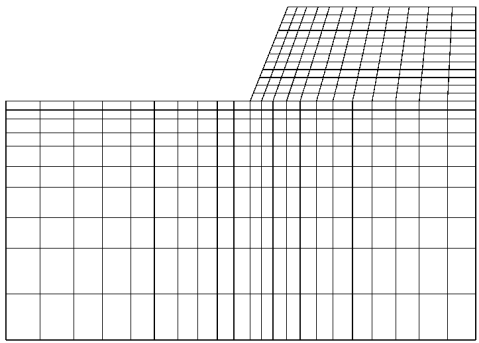}
\vspace*{-5mm}
\caption{\footnotesize Initial mesh.}
\label{fig18}
\end{center}
\end{figure}

The following assumptions have been introduced to simulate the entire forming process:
\begin{itemize}
\item the die is rigid;
\item friction is neglected at the contact between powder and die walls;
\item the initial configuration is that shown meshed in Fig.~\ref{fig18}.
\end{itemize}

After the initial state, defined by given initial values of void ratio 
and confining pressure ($e_0$=2.129 and $p_0$=0.063 MPa have been assumed, respectively), has been 
prescribed, the loading history is assigned, in terms of the following three sequential steps:
\begin{enumerate}
\item forming is prescribed by imposing the motion of the upper part of the boundary
      (3.78~mm, corresponding to the value measured during forming
      at the final load of 50~kN);
\item unloading is simulated by prescribing null forces on the upper part
      of the boundary;
\item ejection is simulated by prescribing null forces on all the boundary.
\end{enumerate}

\begin{figure}[!htb]
\begin{center}
\vspace*{3mm}
\includegraphics[width=8.5cm]{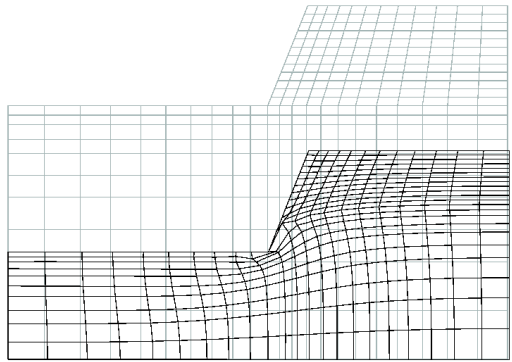}
\vspace*{-5mm}
\caption{\footnotesize Initial and deformed (step~3) meshes.}
\label{fig19}
\end{center}
\end{figure}

The final deformed mesh (at the end of step 3), is reported in Fig. \ref{fig19} superimposed on the initial mesh.
It can be noted from the figure that the elements near the corner of the punch are 
excessively distorted so that results in this zone should be considered unrealistic. 
Comparing the meshes, it can be observed that the deformation suffered by the piece is moderately large. 

\begin{figure}[!p]
\begin{center}
\includegraphics[width=8.5cm]{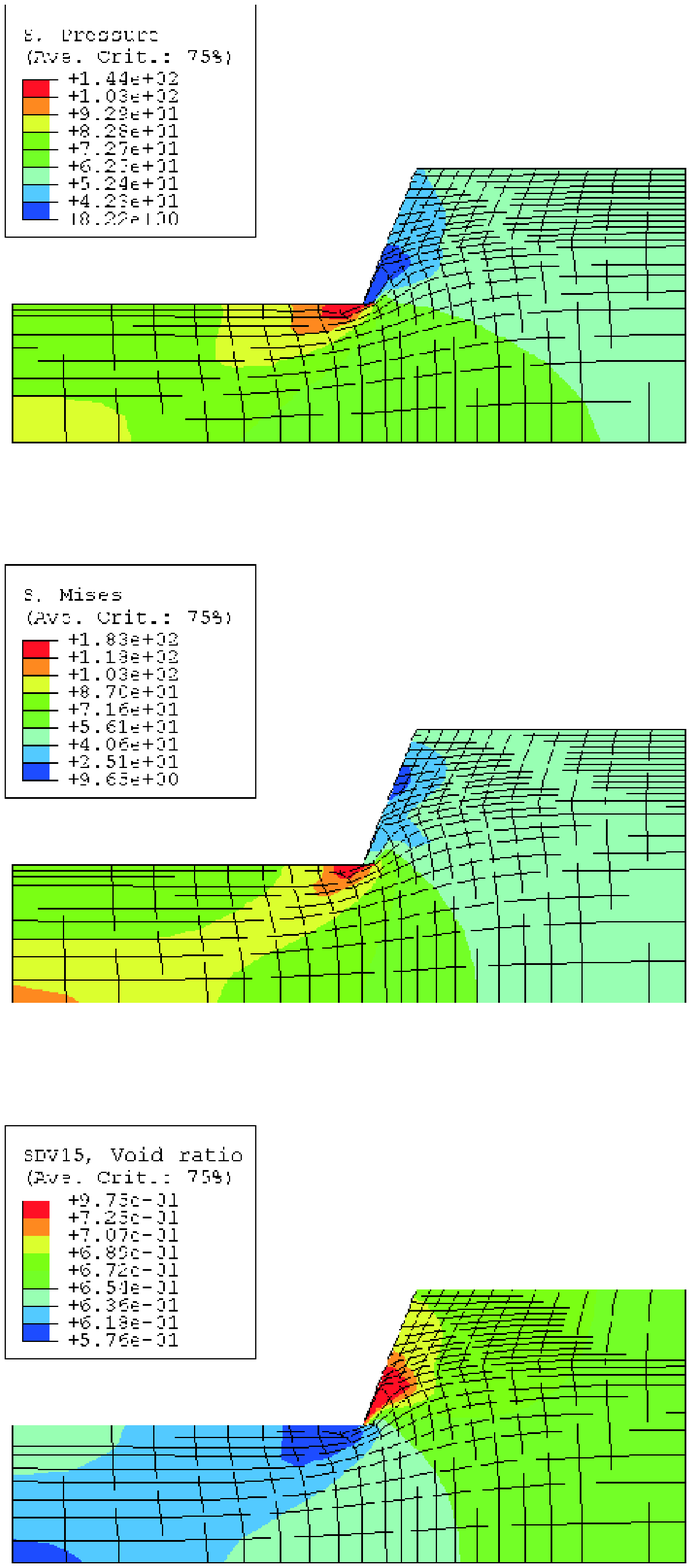}
\caption{\footnotesize Hydrostatic stress $p$ (MPa, upper part), Mises stress $q$ (MPa, central part), 
and Void ratio $e$ (lower part) distributions at the end of step~1 (after loading).}
\label{fig20}
\end{center}
\end{figure}

\begin{figure}[!p]
\begin{center}
\includegraphics[width=8.5cm]{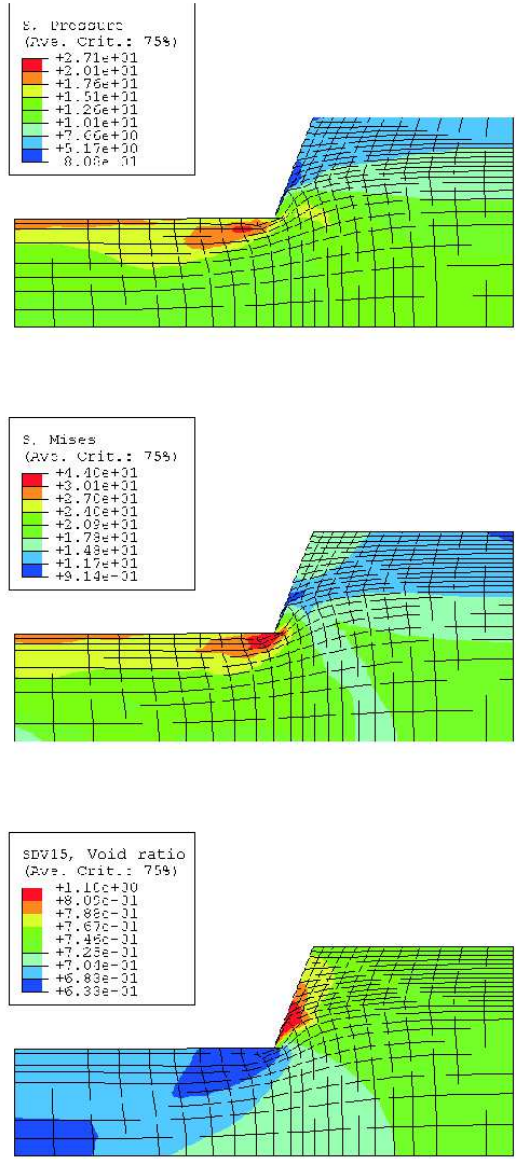}
\caption{\footnotesize Hydrostatic stress $p$ (MPa, upper part), Mises stress $q$ (MPa, central part), 
and Void ratio $e$ (lower part) distributions at the end of step~2 (after punch removal).}
\label{fig21}
\end{center}
\end{figure}

\begin{figure}[!p]
\begin{center}
\includegraphics[width=8.5cm]{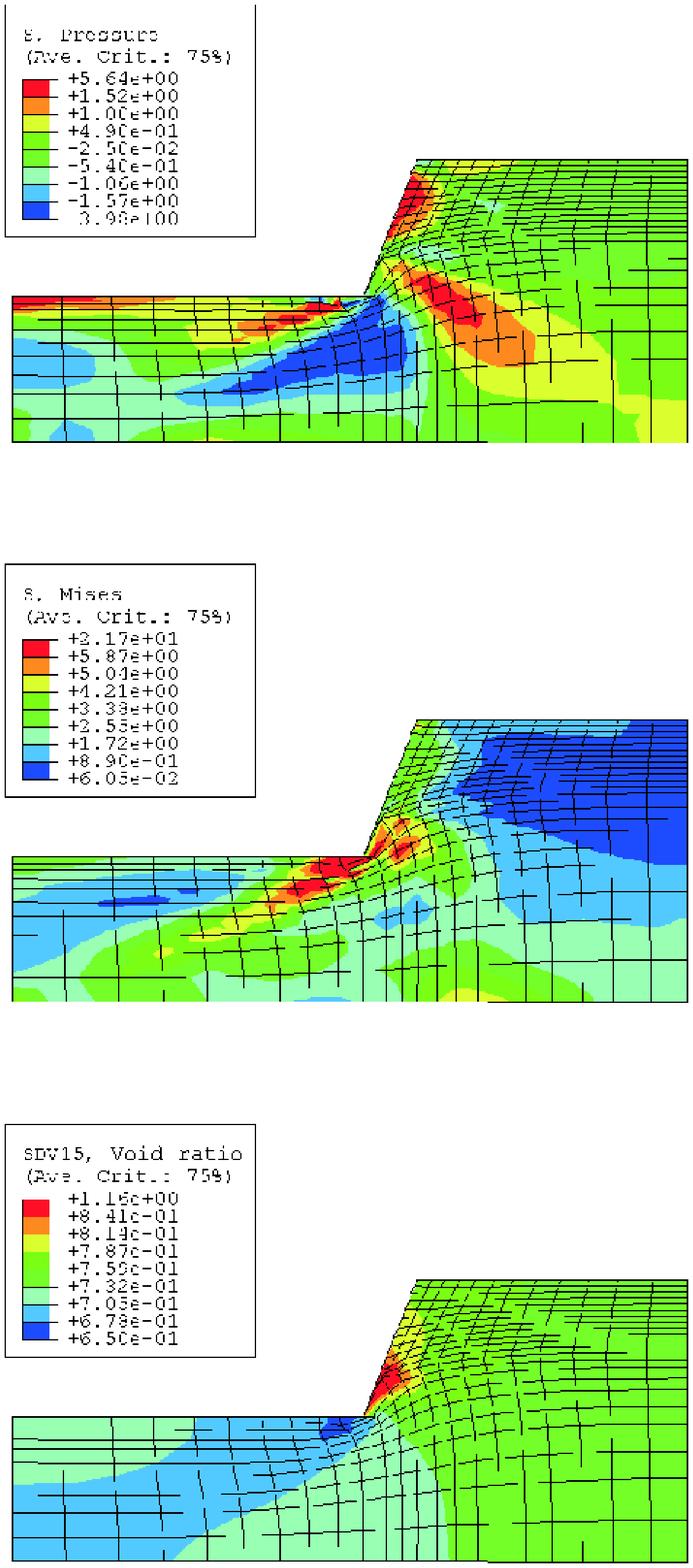}
\caption{\footnotesize Hydrostatic stress $p$ (MPa, upper part), Mises stress $q$ (MPa, central part), 
and Void ratio $e$ (lower part) distributions at the end of step~3 (in the green piece).}
\label{fig22}
\end{center}
\end{figure}

The hydrostatic stress component $p$ (taken positive when compressive, upper part in the figure), 
the Mises stress $q$ (central part in the figure) and the void ratio 
$e$ (lower part in the figure) are reported in Fig.~\ref{fig20} at the end of step~1, in Fig.~\ref{fig21} at the end of step 2, and 
in Fig. \ref{fig22} at the end of step 3.

Regarding Fig. \ref{fig20} (representative of step 1) we may note that,
excluding the small, unrepresentative zone near the corner of the punch, the hydrostatic stress $p$ ranges 
between 
40 and 100~MPa and the Mises stress $q$ between 25 and 120~MPa, evidencing a high stress inhomogeneity. 

Considering Fig. \ref{fig21} (representative of step 2) 
it may be important to note that residual stress is definitely high, due to the lateral constraint 
(provided by the walls of the mold) still present at 
the end of step~2. The knowledge of the lateral stress is important for practical purposes since 
the force needed 
for the ejection of the final piece can be estimated from this value employing the Coulomb friction law. 
In particular, a rough, but simple evaluation can be immediately obtained from numerical output at the 
end of step~2 employing the formula
\begin{quote}
\hspace*{-4mm}
{\it ejection~force} = $\alpha\, \tan\, \phi$
({\it mean~lateral~stress $\times$ lateral~surface~of~the~piece}),
\end{quote}
where $\phi$ is the powder friction angle (equal to 32$^\circ$ in our case) and $\alpha$ is a coefficient 
dependent on the roughness of the die wall and ranging between 0 and 1.

Regarding Fig. \ref{fig22} (representative of step 3), we note that $p$ and $q$ represent, in 
terms of hydrostatic stress and Mises stress components, the residual stress 
distribution in the green body at the end of forming.

Excluding again the unrepresentative zone near the corner of the punch, the hydrostatic stress $p$ ranges now 
between -1.5 and 6~MPa and the Mises stress $q$ between 1 and 6~MPa. Moreover, the void ratio 
varies between 0.6 and 0.9. It can be noted that the minimum void ratio is not associated with the maximum 
residual mean stress, it is rather associated with the maximum mean stress reached during loading (step 1). The 
results suggest also that two oblique zones of material are formed, the outer of which is subject to high compressive 
mean stresses, whereas the inner is subject to tensile stresses, creating a sort of truss-like mechanism. 
This can be considered representative of a 
situation where the tensile stresses tend to open possible microcracks, leading to defects formation 
in the green. It is however worth remembering that, even in the absence of macro defects, 
the mechanical behaviour of the green and the 
shrinkage during future sintering are deeply affected by the inhomogeneities in the residual stress and density 
distributions.

From the comparison reported in Fig.~\ref{fig19}
between the initial mesh and that at the end of step 3,
it becomes now possible to evaluate the springback effect. In contrast to the prediction of the 
simple model employed by Piccolroaz et al. (2002), it can be noted that now the model correctly 
predicts that the springback effect and the shape distortion are very small. In particular, the final diameter 
of the piece is estimated to be 0.1~mm larger than the inner diameter of the die, in agreement 
with our experimental observations.

\begin{figure}[!p]
\begin{center}
\includegraphics[width=8.5cm]{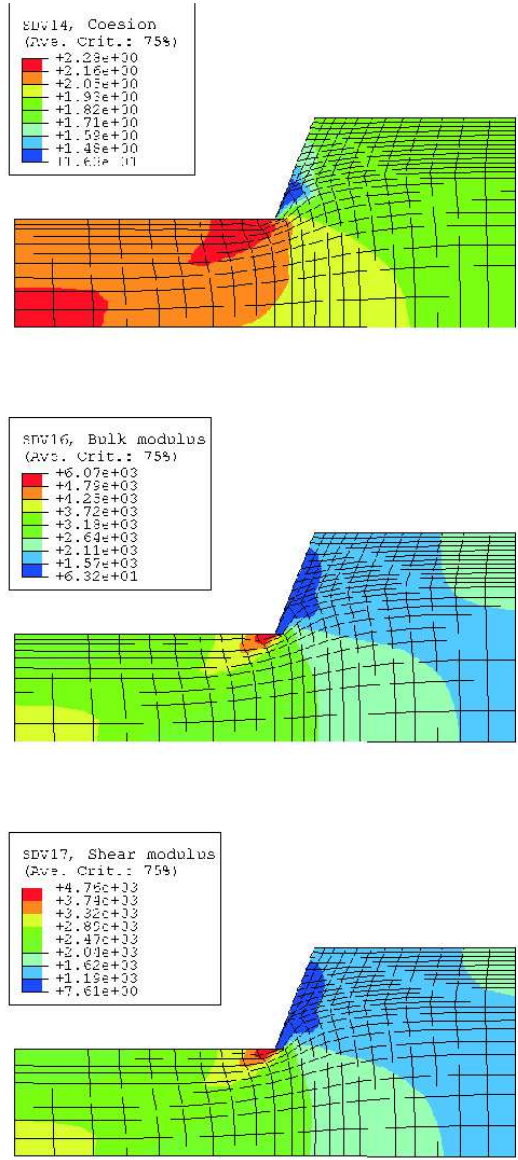}
\caption{\footnotesize Distribution of cohesion $c$ (MPa, upper part), bulk modulus $K_t$ (MPa, central 
part), and shear modulus $\mu$ (MPa, lower part) at the end of step~3 (in the green piece).}
\label{fig23}
\end{center}
\end{figure}

The cohesion $c$ attained by the material at the end of the overall process is shown in Fig.~\ref{fig23}, 
upper part, whereas 
the elastic properties of the final piece are reported in Fig.~\ref{fig23} 
(central and lower parts, respectively), in 
terms of tangent 
bulk modulus $K_t$ and shear modulus $\mu$. 
The inhomogeneity of the elastic properties evidenced in Fig.~\ref{fig23} is obviously a
consequence of elastoplastic coupling. This effect and also the increase in cohesion 
have been not modelled in the simple analysis presented by Piccolroaz et al. (2002).
It may be observed by comparing the maps shown in Fig.~\ref{fig23} with the map of hydrostatic stress $p$ 
at the end of step 1 (Fig.~\ref{fig20}, upper part) 
that there is a strong relation between mechanical properties gained 
by the material in the 
final piece and the mean stress reached during loading. This results from our 
analyses to represent the most important parameter in the entire forming process.

\begin{figure}[!htb]
\begin{center}
\vspace*{3mm}
\includegraphics[width=7.5cm]{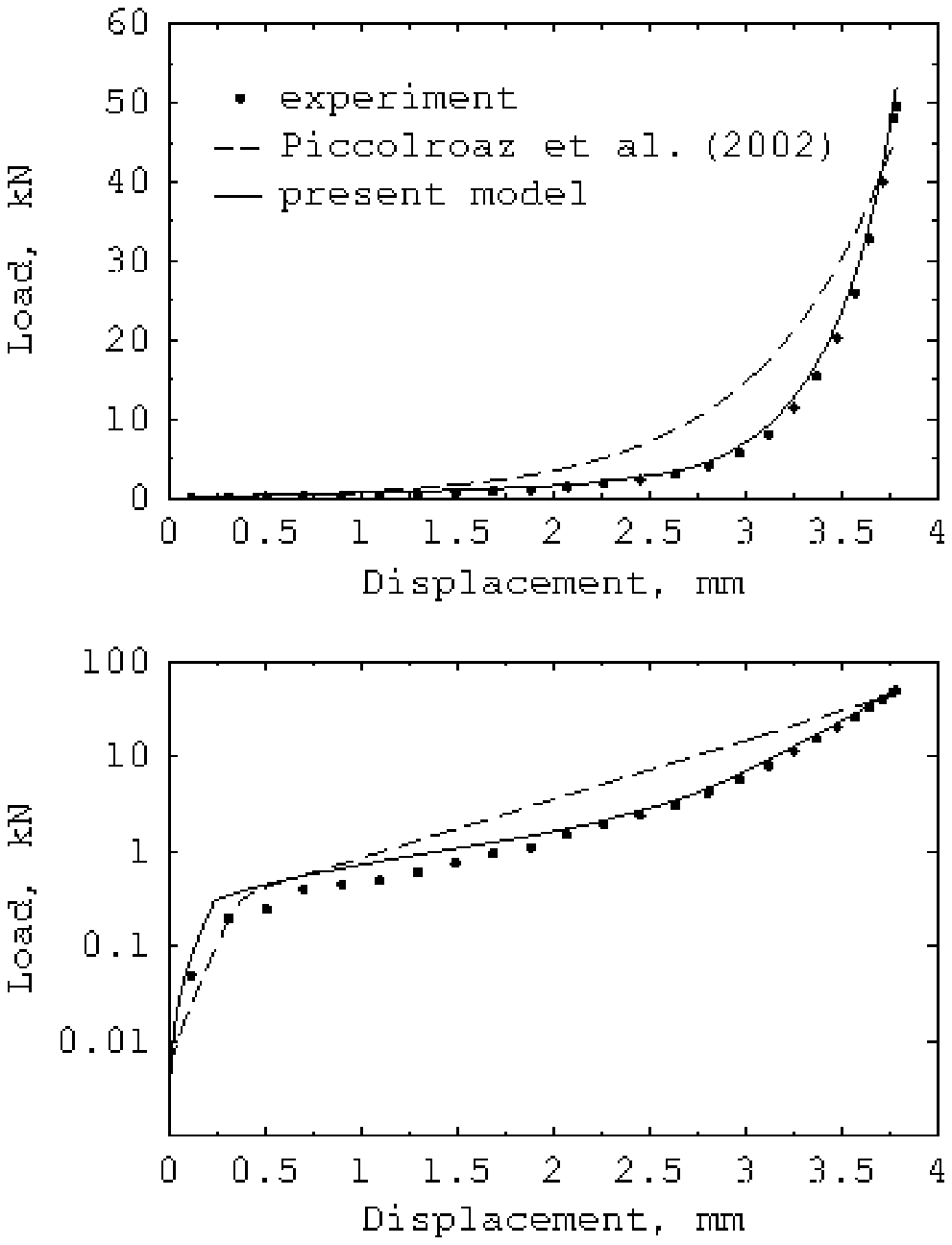}
\caption{\footnotesize Experimental (M KMS-96 alumina powder) and simulated load vs.\ displacement curves during
forming of the piece shown in Fig. \ref{fig17}, in natural and semilog representations. 
Results by Piccolroaz et al. (2002) are also reported (dashed).}
\label{fig24}
\end{center}
\end{figure}

Experimental and simulated load displacement curves during forming (natural and semilogarithmic representations are reported)
are compared in Fig.~\ref{fig24}, together with the results obtained by Piccolroaz et al. (2002) included
(dashed) in the figure.
Beyond the excellent agreement, it may be instructive to compare the present numerical simulation with that 
performed by Piccolroaz et al. (2002) employing a simple model. First, we note that the simple
model approximately describes two straight lines, whereas the present model describes a curved line, much 
closer to the experimental results. Second, we may speculate on the limits of the approach presented 
by Piccolroaz et al. (2002) and of many similar models available in the literature; 
in particular, the modelling can be accurate enough, if intended 
to predict a `global' force-displacement curve like that reported in Fig.~\ref{fig24}.
On the other hand, the simple model does not predict increase in cohesion 
and dependence of elastic properties on
plastic deformation, so that the internal stress and strain distributions result almost completely different.

\begin{figure}[!htb]
\begin{center}
\vspace*{3mm}
\includegraphics[width=7cm]{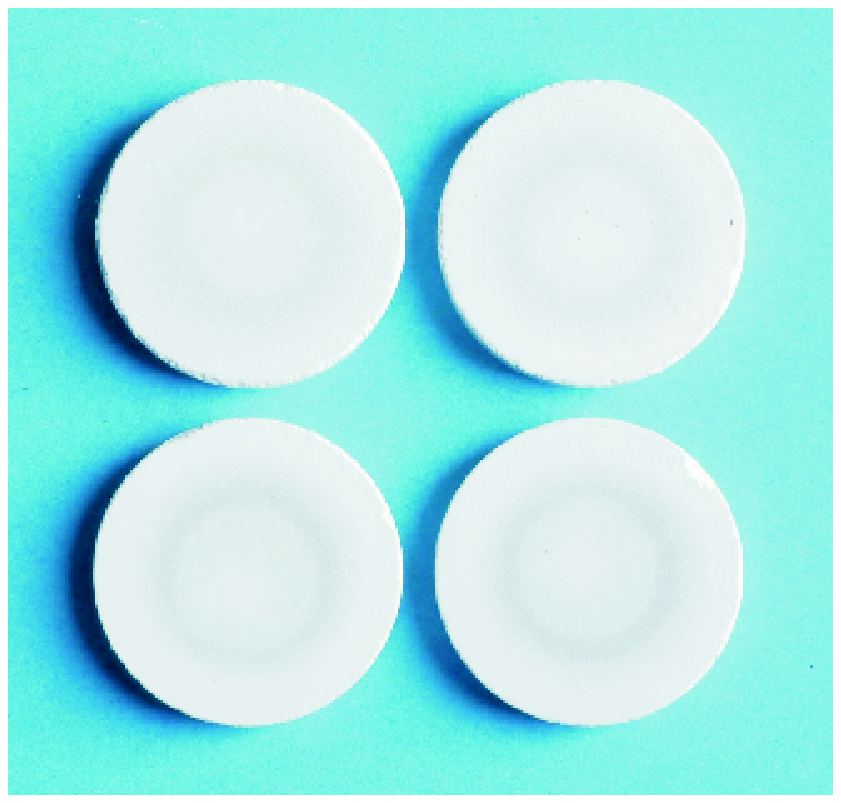}
\caption{\footnotesize Photograph of the lower side of the formed pieces.}
\label{fig25}
\end{center}
\end{figure}

\begin{figure}[!htb]
\begin{center}
\vspace*{3mm}
\includegraphics[width=7cm]{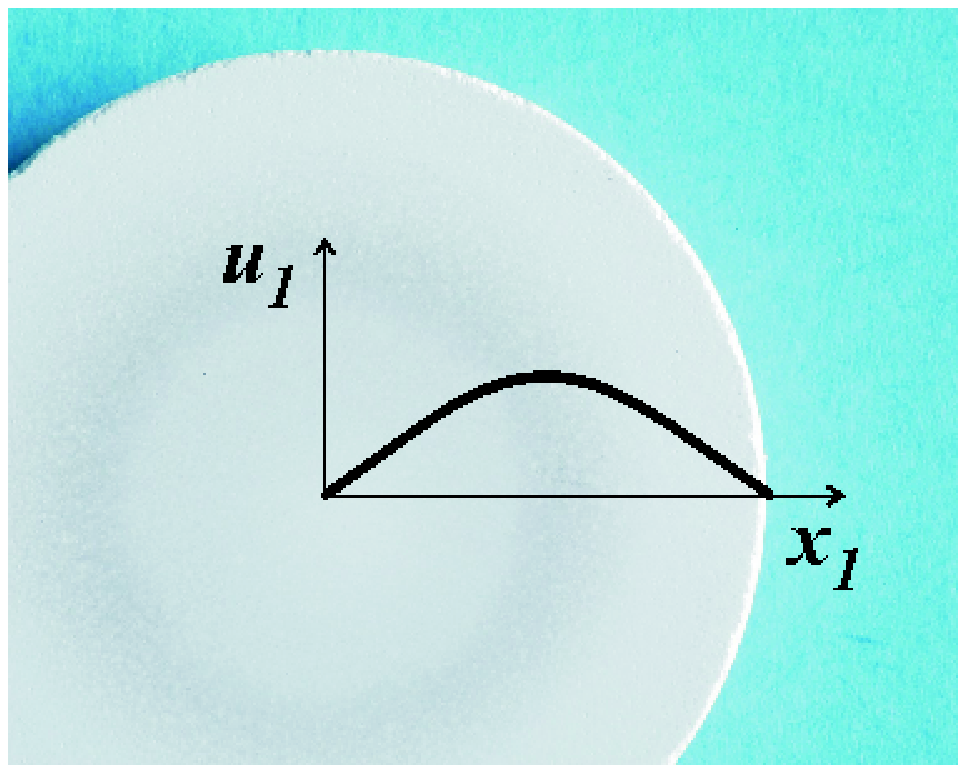}
\caption{\footnotesize Simulated radial displacements superimposed to the photograph of the 
lower surface of the formed piece.}
\label{fig26}
\end{center}
\end{figure}

A photograph of the lower side of the formed pieces is shown in Fig.~\ref{fig25}, where we can note the formation 
of annular dark zones, evidencing a sliding between material and mold. 
This sliding is indeed predicted by the 
simulations, so that the radial displacement at the mold contact is reported in Fig.~\ref{fig26}, superimposed 
to the photograph of one of the pieces shown in Fig.~\ref{fig25}, so that, since 
the dark zone corresponds to the peak of the radial displacement, simulations 
again agree with experimental observations.

\subsection{The effects of large strains}

In order to have an insight on the possible effects of large strains, a few analyses have been 
performed using the option NLGEOM available on ABAQUS, still employing the presented small-strain formulation. 
\begin{figure}[!htb]
\begin{center}
\vspace*{3mm}
\includegraphics[width=7cm]{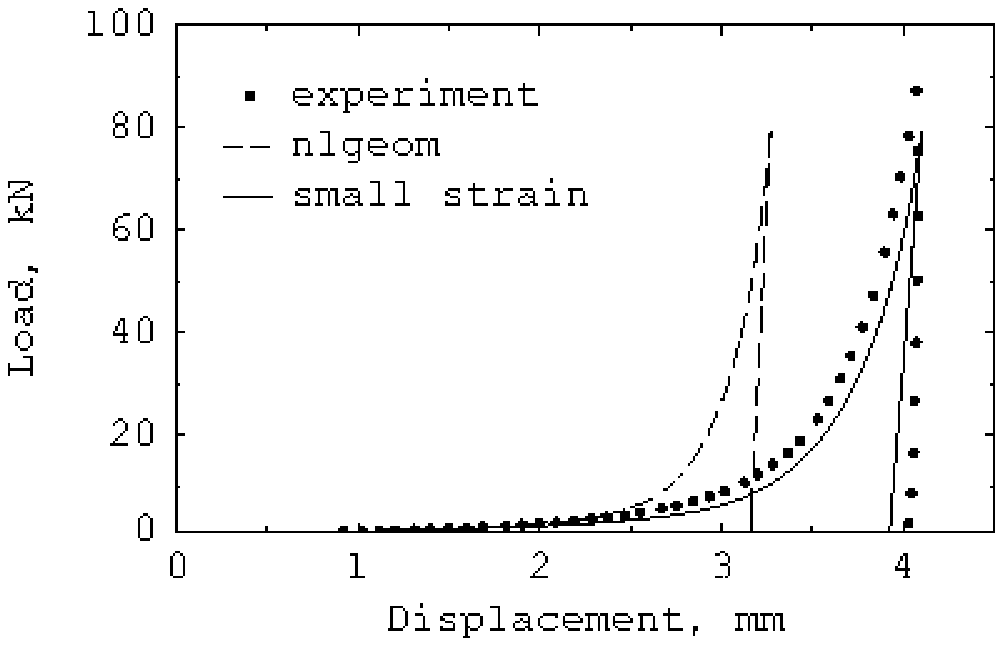}
\caption{\footnotesize Experimental (M KMS-96 alumina powder) and simulated load vs.\ displacement curves
for uniaxial strain. The prediction obtained employing the option NLGEOM has been reported (dashed) together 
with the prediction relative to the small strain assumption.}
\label{fig27}
\end{center}
\end{figure}
A representative result is shown dashed in Fig.~\ref{fig27}, relative to the forming of a tablet 
[see Fig. \ref{fig15} (d)]. Obviously, {\it a rigorous analysis would require a complete 
model reformulation and reinterpretation of experimental results}. 
However, we believe that due to the fact that the deformations are only moderately large
and do not involve rotations, the simulation 
shown in Fig.~\ref{fig27} should at least give some understanding of the main 
differencies between predictions expected from the two formulations.
We note that the effect of geometrical nonlinearities yields, as expected, a stiffening of the 
response, but does not change the results qualitatively. 
However, the quantitative difference may be enough to suggest the interest in a large strain formulation, 
which is given in the Part II of this paper (Piccolroaz et al. 2005).

\section{Conclusions}

Results presented of this paper provide a constitutive framework to realistically describe 
forming processes of ceramic materials. Even if the experimental results are still 
incomplete and the employed elastoplastic model has been developed 
in a small strain formulation, it has been shown that it is possible to predict:
\begin{itemize}
\item the force needed for mold ejection,
\item the springback effect and related shape distortion of formed pieces,
\item the residual stress distribution,
\item the gain in cohesion and the final elastic properties,
\item the density distribution and the related presence of defects in the green body.
\end{itemize}
The last of the above points is related to the prediction of defects in 
the sintered piece and therefore its investigation has an important consequences in the design of 
ceramic elements. 

In closure, we mention that the present modelling can be extended in different directions. 
Introducing thermoplastic effects, the sintering phase might be covered by modelling, so that simulation could be 
extended to the entire production process. Moreover, both sintering aids and powder characteristics might enter 
the elastic-plastic constitutive laws, so that the optimal powder composition and morphology could be predicted 
for different forming problems. 

\vspace*{3mm}

\vspace*{3mm}
\noindent
{\sl Acknowledgments }

\noindent
\noindent 
Financial support of 
MIUR-COFIN 2003 `Fenomeni di degrado meccanico di interfacce
in sistemi strutturali: applicazioni in Ingegneria Civile ed a campi di ricerca
emergenti' is gratefully acknowledged.

\newpage

{
\singlespace

}

\clearpage
\setcounter{equation}{0}
\renewcommand{\theequation}{{A}.\arabic{equation}}
\begin{center}
{\bf APPENDIX A. Powder characteristics}\\
\end{center}

Calibration of the model has been performed on the basis of experiments both already available and carried out 
on a commercial ready-to-press alumina powder (96\% purity), manufactured by
Martinswerk GmbH (Bergheim, Germany) and identified as 392 Martoxid KMS-96. The data
presented by the manufacturer are given in Table~\ref{tab1-pic}. It can be noted from
the upper part of Fig.\,\ref{fig01} that the granules have a mean diameter of 250\,$\mu$m.
\begin{table}[!htb]
  \footnotesize
  \vspace*{2mm}
\begin{center}
  \caption{Granulometric and density properties of the tested alumina powder.}
  \label{tab1-pic}
\begin{tabular}{||l|c||}
\hline\hline
MWM 28 Vibration sieving                                      & ~        \\
sieve residue $>$ 300\,$\mu$m                                 & 3.9\%    \\
sieve residue $>$ 150\,$\mu$m                                 & 56.3\%   \\
sieve residue $<$ 63\,$\mu$m                                  & 2.5\%    \\ \hline
Bulk density~~(g/cm$^3$)                                      & 1.219    \\ \hline
Green density (p = 50\,MPa)~~(g/cm$^3$)                       & 2.39     \\ \hline
Fired density (T=1600$^\circ$C, 2h)~~(g/cm$^3$)               & 3.77     \\ \hline
\hline
\end{tabular}
\end{center}
\end{table}

\clearpage
\setcounter{equation}{0}
\renewcommand{\theequation}{{B}.\arabic{equation}}
\begin{center}
{\bf APPENDIX B. Yield function gradient}\\
\end{center}

The gradient $\bQ$ of the yield function (\ref{yieldfunction}) is [a detailed derivation
can be found in (Bigoni and Piccolroaz, 2004)]
\beq
\lb{grad}
\bQ = \frac{\partial F}{\partial \bsigma} = 
a(p)\, \Id + b(\theta)\, \tilde{\bS} + c(\theta)\, \tilde{\bS}^{\perp},
\eeq
where
\beq
\tilde{\bS} = \sqrt{\frac{3}{2}}\frac{\dev \bsigma}{q},~~~~
\tilde{\bS}^{\perp} = -\frac{\sqrt{2}}{\sqrt{3}\, q} \frac{\partial \theta}{\partial \bsigma} = 
\frac{1}{\sin 3 \theta}\left[\sqrt{6}\left(\tilde{\bS}^{2}-\frac{1}{3} \Id\right)
-\cos 3\theta\, \tilde{\bS} \right],
\eeq
and
\beqar
\lb{abc}
a(p) &=& -\frac{1}{3}\deriv{f(p)}{p} = \frac{Mp_c}{3(p_c+c)}
                                  \frac{\left(1-m\Phi^{m-1}\right)
                                  \left[2(1-\alpha)\Phi+\alpha\right]+
                                  2(1-\alpha)\left(\Phi-\Phi^m\right)}
                                  {2\sqrt{\left(\Phi-\Phi^m\right)
                                  \left[2(1-\alpha)\Phi+\alpha\right]}}, \nonumber \\
b(\theta) &=& \sqrt{\frac{3}{2}} \frac{1}{g(\theta)}, \\
c(\theta) &=& -\frac{\sqrt{3}\gamma\sin 3 \theta}{\sqrt{2}\sqrt{1-\gamma^{2}\cos^{2} 3\theta}}
\sin\left[\beta\frac{\pi}{6}-\frac{1}{3}\cos^{-1}(\gamma\cos 3\theta)\right]. \nonumber
\eeqar

\clearpage
\setcounter{equation}{0}
\renewcommand{\theequation}{{C}.\arabic{equation}}

\end{document}